\documentclass[twocolumn,amssymb,pra,superscriptaddress]{revtex4}

\usepackage{graphicx}
\usepackage{multirow}
\usepackage{amsmath, amssymb}
\usepackage{subfigure}
\usepackage{rotating}

\usepackage[normalem]{ulem} 
\usepackage{color} 

\begin{document}

\newcommand{\todo}[1]{\textcolor{green}{\textit{\textbf{#1}}}}
\newcommand{\toadd}[1]{\textcolor{blue}{\uwave{#1}}}
\newcommand{\tosub}[1]{\textcolor{red}{\sout{#1}}}

\newcommand{\thecolwidth}{85mm}

\newcommand{\placefigone}{\begin{figure}
  \centering
  \includegraphics[width=\thecolwidth]{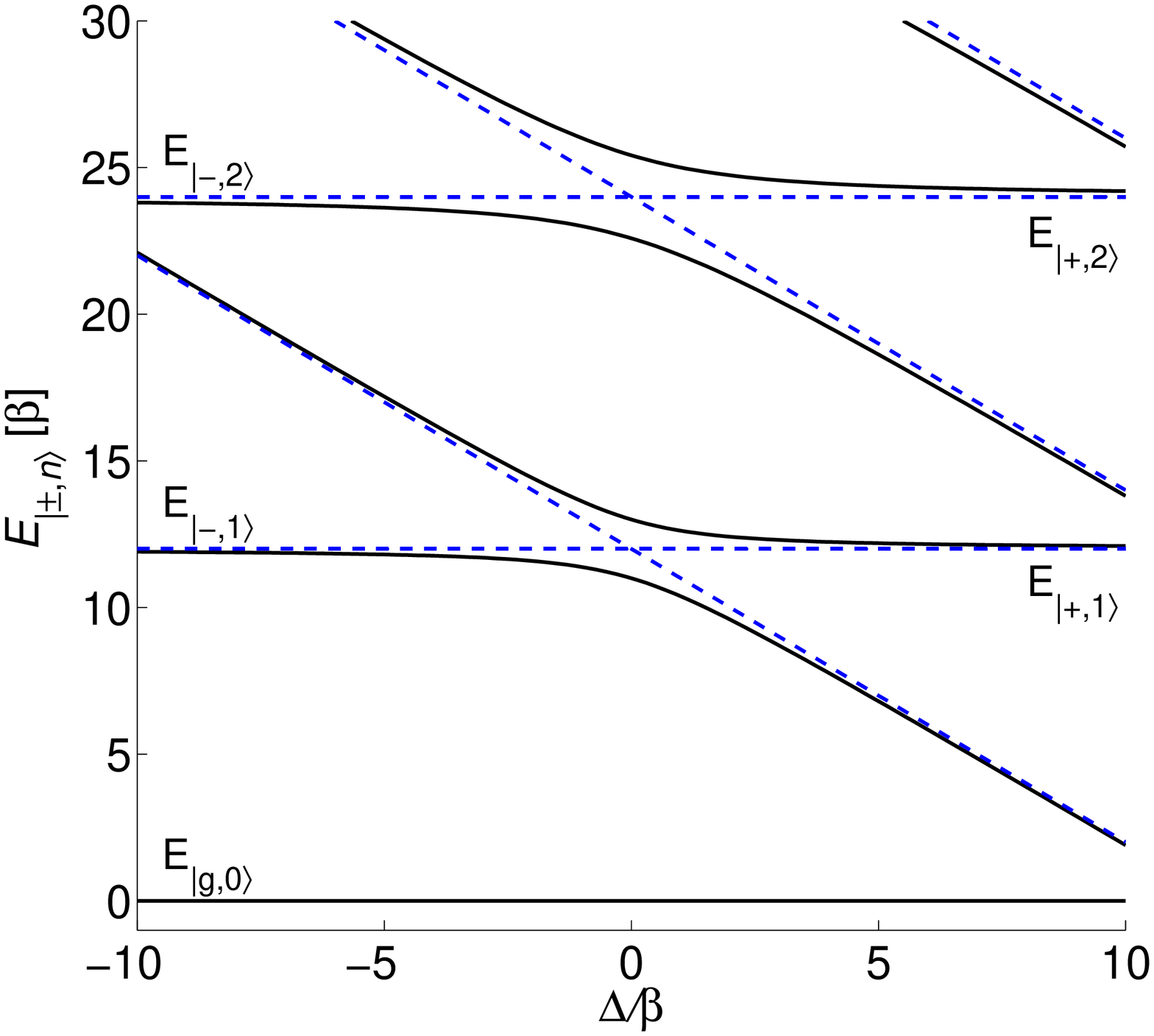}
  \caption{(Color online) Energy eigenvalues
    [Eq.~(\ref{eq:singlecavityeigenvalues})] for a single cavity as a
    function of the detuning $\Delta$.  We set $\omega=12\beta$ for
    illustrative purposes.  The blue dashed lines indicate the
    asymptotes of the energies for each band, the diagonal lines
    represent the energy of the cavity due to the atom, while the
    horizontal lines represent the energy due to the photons only.}
  \label{fig:JCEig}
\end{figure}}

\newcommand{\placeimplementation}{\begin{figure}[tbp]
  \centering
  \includegraphics[width=\thecolwidth]{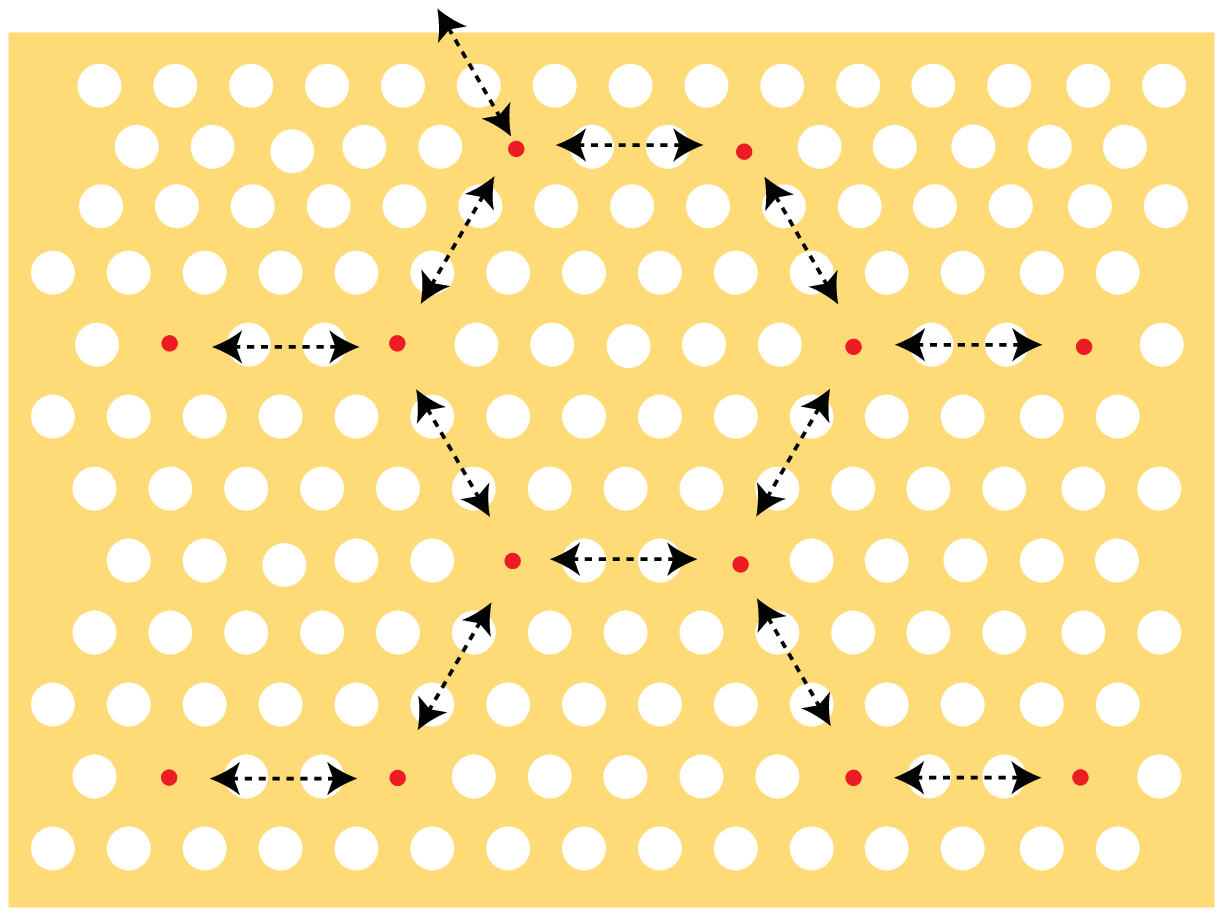}
  \caption{(Color online) This schematic shows a possible
    implementation of the system, for three nearest neighbors.  The
    dielectric medium is shown in yellow.  The photonic crystal is
    made by periodic variations in refractive index, caused by
    drilling holes (white disks).  The cavities are regions where
    holes have not been drilled - effectively, the undrilled holes.
    In each cavity, a red disk represents the two-level atom.  The
    arrows indicate the nearest neighbors, across which photons can
    tunnel with hopping rate $\kappa$.}
  \label{fig:implementation}
\end{figure}}

\newcommand{\placeenergies}{\begin{figure}
  \centering
  \includegraphics[width=\thecolwidth]{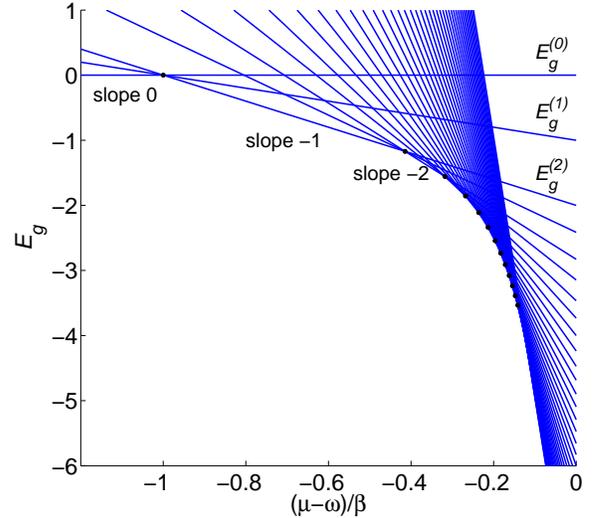}
  \caption{(Color online) Ground state energies for each block of the
    Hamiltonian, Eq.~(\ref{eq:mainham}), as a function of
    $(\mu-\omega)/\beta$, with $\kappa=0$.  Recall that
    $\langle\hat{L}\rangle$ is equal to the negative of the slope
    (with respect to $\mu$) of the smallest energy eigenvalue of the
    \emph{whole} Hamiltonian, and that this slope is always an
    integer.  This figure therefore shows explicitly how the phase
    diagrams are constructed - the black points indicate the
    boundaries between plateaus.}
  \label{fig:energies}
\end{figure}}

\newcommand{\placeexact}{\begin{figure}[h!tbp!]
  \begin{center}
    \includegraphics[width=\thecolwidth]{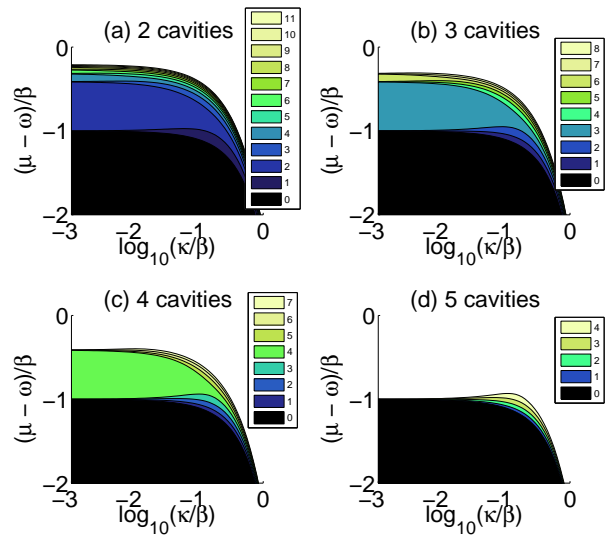}
    \caption{(Color online) These plots show the expectation value of
      the total number of excitations $\langle \hat{L}\rangle$ as a
      function of $(\mu-\omega)/\beta$ and $\kappa/\beta$, for two,
      three, four and five cavities in periodic boundary conditions,
      with $\Delta=0$.  Note that the top boundary in each plot is the
      limit of calculations.  The Hamiltonian matrix used to create
      each plot is truncated at $l=12, 9, 8$ and 5, respectively.}
    \label{fig:exact}
  \end{center}
\end{figure}}

\newcommand{\placetwocavsExplain}{\begin{figure}
  \centering
  \includegraphics[width=\thecolwidth]{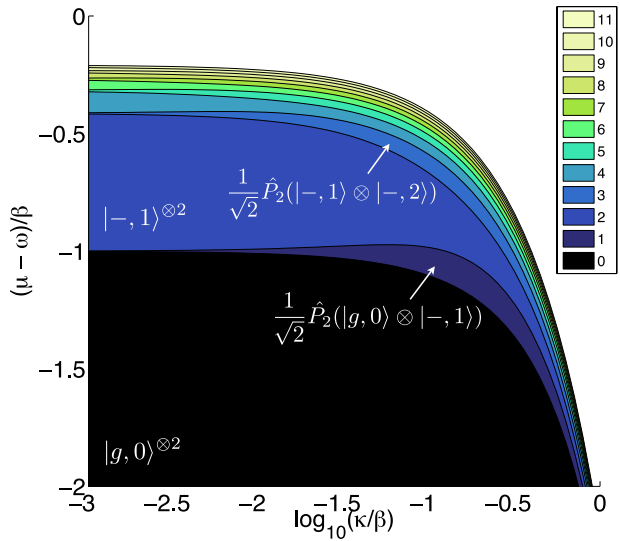}
  \caption{(Color online) The expectation value of the total number of
    excitations $\langle\hat{L}\rangle$ of two cavities with periodic
    boundary conditions in the ground state.  The eigenstates of the
    four lowest plateaus are marked, (see Table \ref{tab:states}).
    The upper boundary marks the limit of calculations.}
  \label{fig:2cavsExplain}
\end{figure}}

\newcommand{\placeofDelta}{\begin{figure}[t!]  \centering
    \subfigure[$\kappa/\beta=0$]{\includegraphics[width=\thecolwidth]{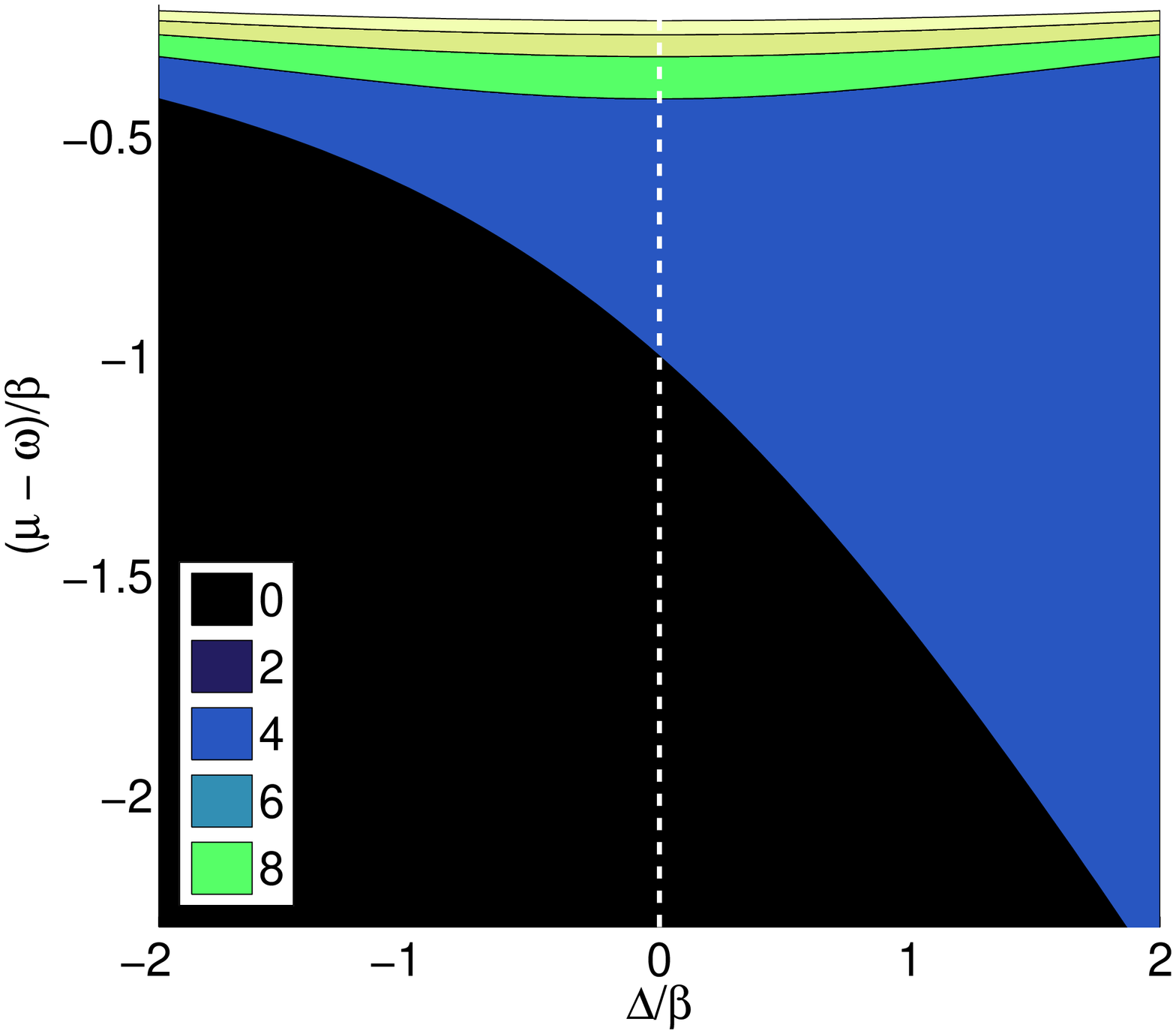}}
    \subfigure[$\kappa/\beta=10^{-1/2}$]{\includegraphics[width=\thecolwidth]{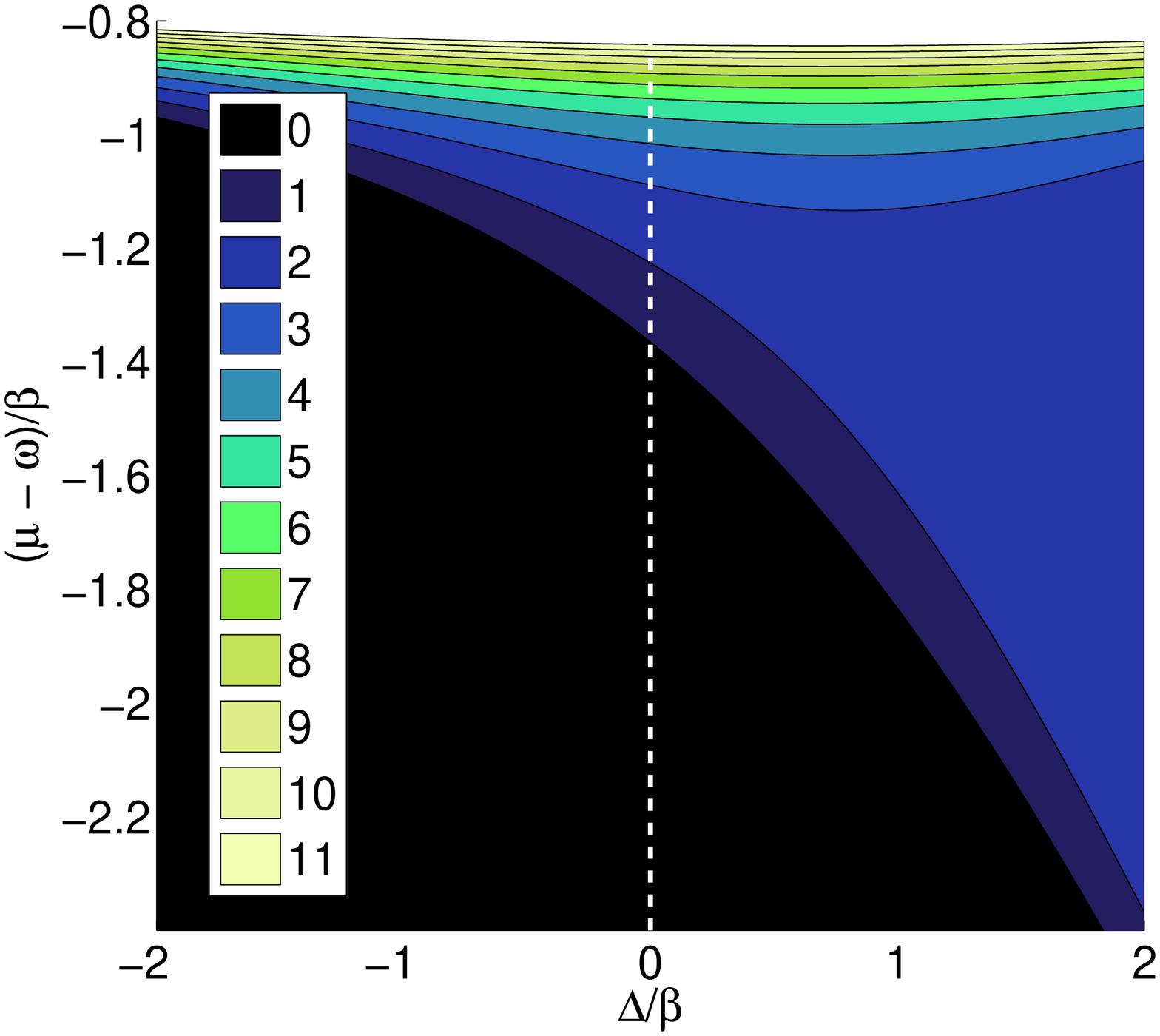}}
    \caption{(Color online) These plots show $\langle \hat{L}\rangle$
      as a function of $(\mu-\omega)/\beta$ and $\Delta/\beta$ for two
      cavities in periodic boundary conditions with (a) $\kappa=0$ and
      (b) $\kappa=10^{-1/2}\beta$.  The upper boundary in both cases
      marks the limit of calculations.  Note that only even plateaus
      are present in (a); this is because of the pinching effect as
      $\kappa\rightarrow0$ - in this limit, plateaus corresponding to
      fractional occupation do not exist.  The white dashed line marks
      $\Delta/\beta=0$, and aids the eye in seeing that the boundaries
      above the first (second) are symmetric (asymmetric) in plot (a)
      [(b)].}
\label{fig:ofDelta}
\end{figure}}

\newcommand{\placematches}{\begin{figure}[h!]
  \begin{center}
    \includegraphics[width=\thecolwidth]{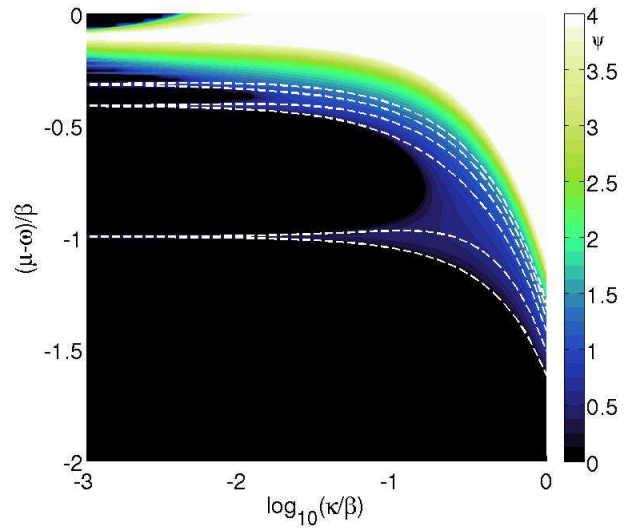}
    \caption{(Color online) This plot shows the mean-field result for
      $z=1$, overlaid with the exact cavity results for two cavities
      with one connection (white dashed lines), as in the fifth row of
      Table \ref{tab:states}.  Note the excellent agreement between
      exact results and mean-field approximation.}
    \label{fig:matches}
  \end{center}
\end{figure}}

\newcommand{\placesuperfill}{\begin{figure}[h!]  \centering
    \includegraphics[width=\thecolwidth]{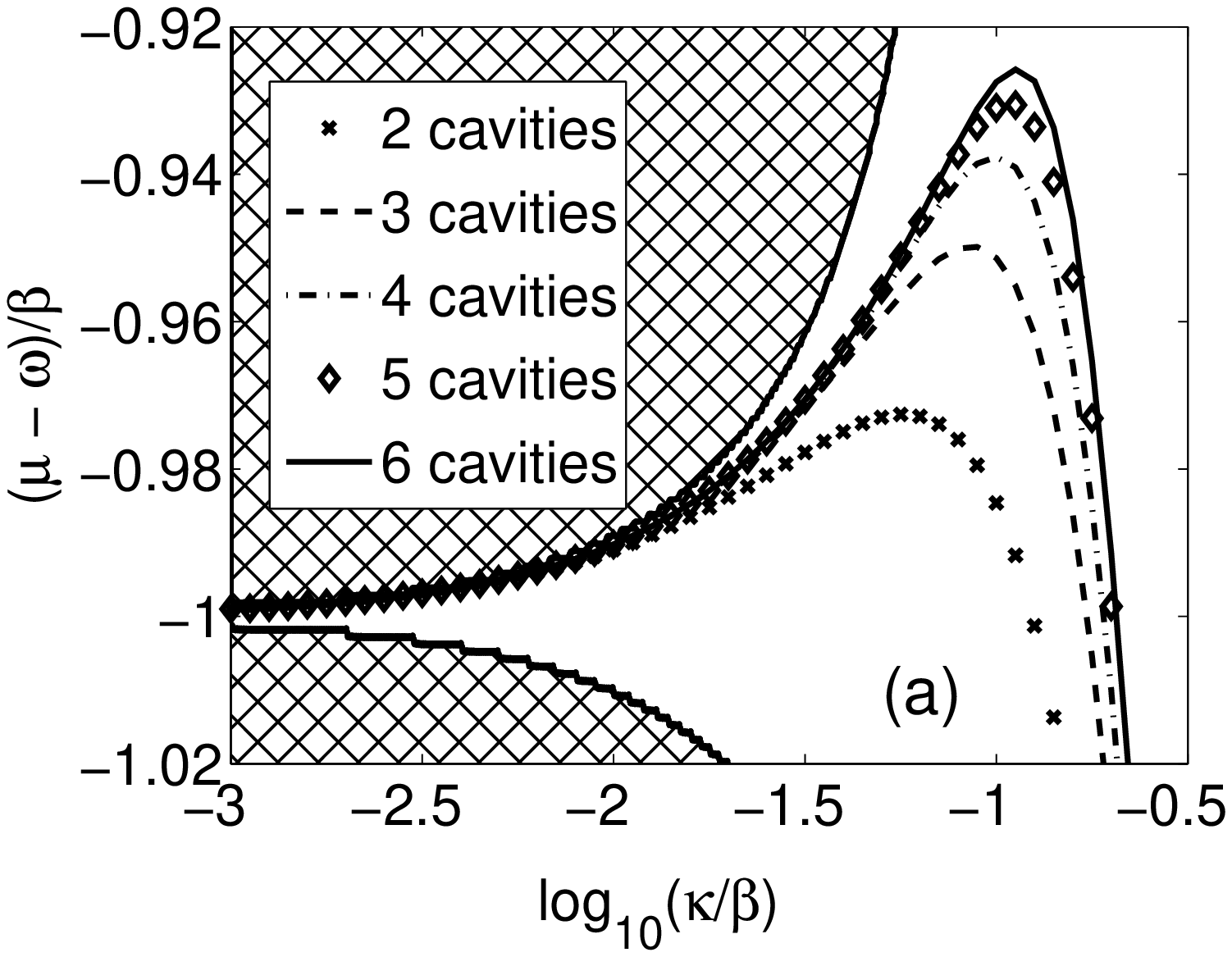}\\
    \includegraphics[width=\thecolwidth]{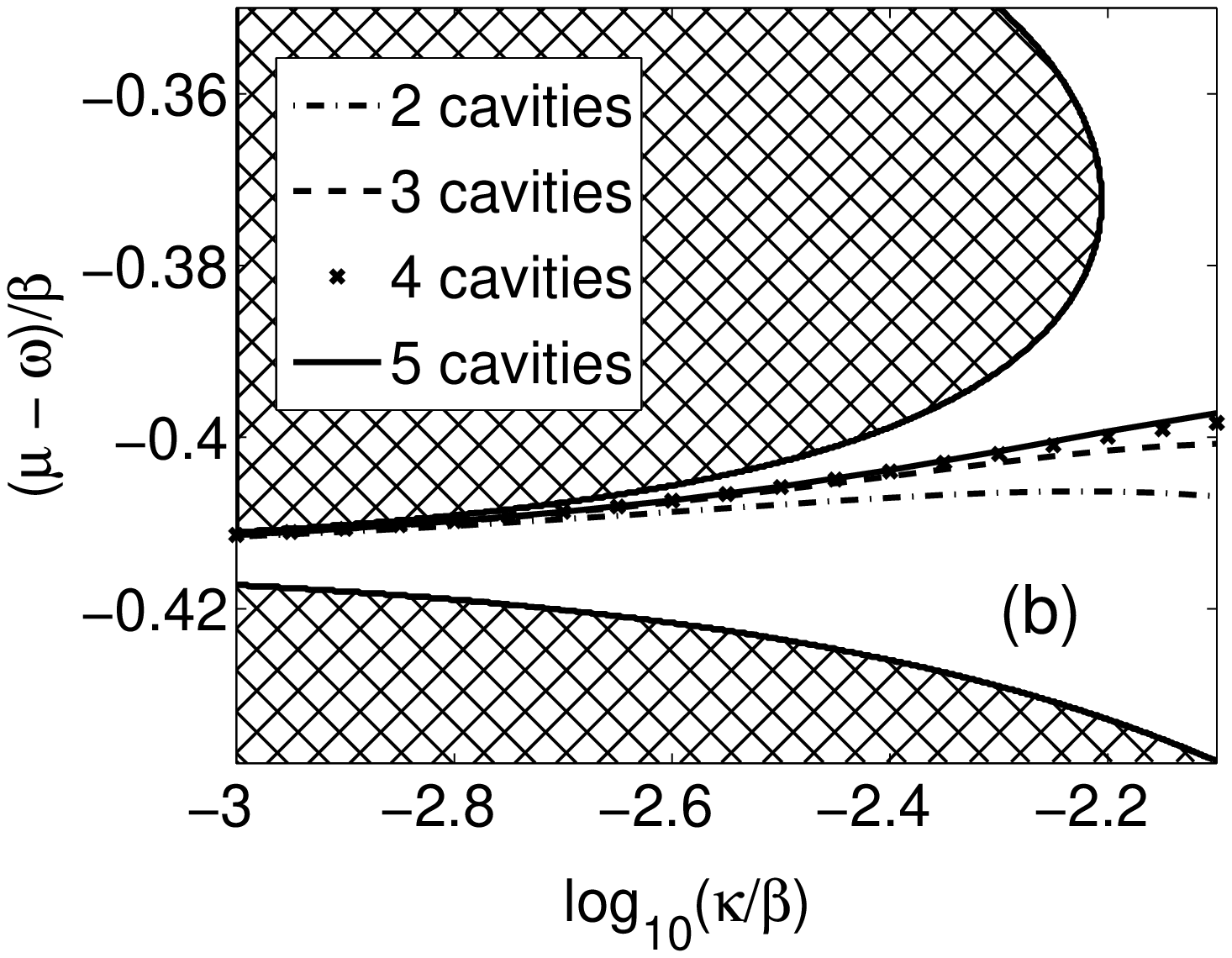}
    \caption{The hatched region represents the $\psi=0$ area of the
      mean-field approximation for two nearest neighbors.  The lines
      (or markers, for distinction purposes only) represent the exact
      calculations.  (a) shows the boundary between the zeroth and
      first lobe, and (b) shows the boundary between the first and
      second lobe. Note that as the number of cavities increases, the
      lines tend to hug the upper mean-field lobe.  This indicates
      qualitatively that as $N\rightarrow\infty$, the exact
      calculations should approach the mean-field ($N=\infty$) limit.}
  \label{fig:superfill}
\end{figure}}

\newcommand{\placetemperatureanddisorders}{\begin{figure}[t]
  \centering
  \subfigure[]{\includegraphics[width=\thecolwidth]{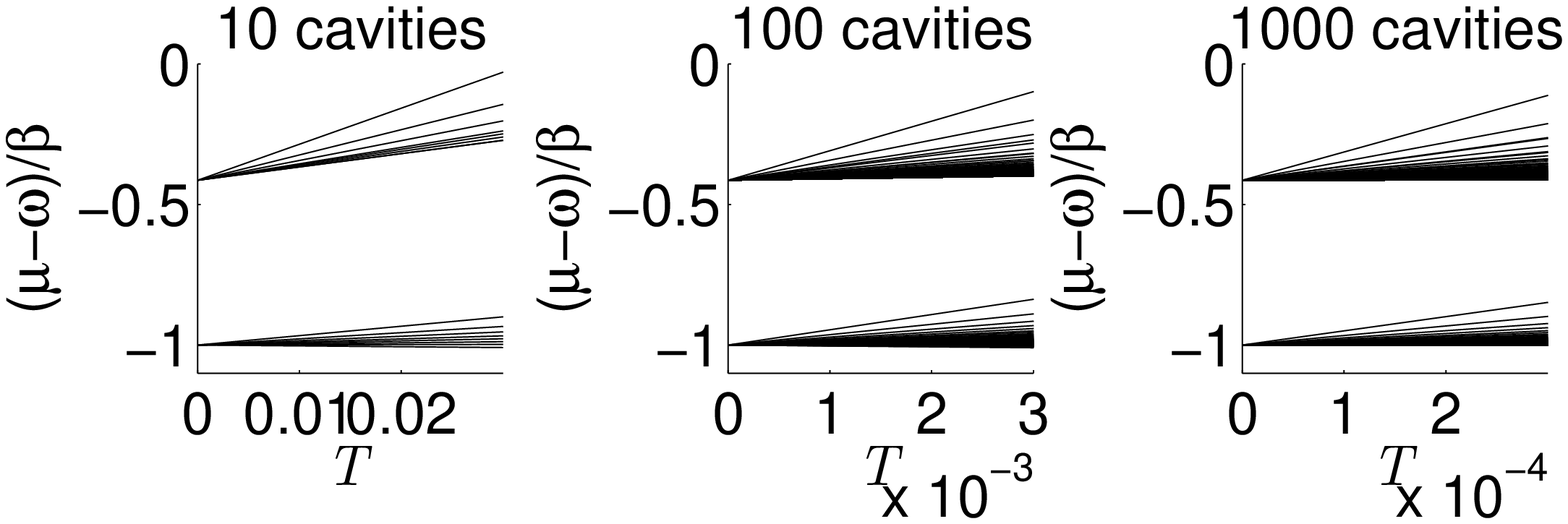}}
  \subfigure[]{\includegraphics[width=\thecolwidth]{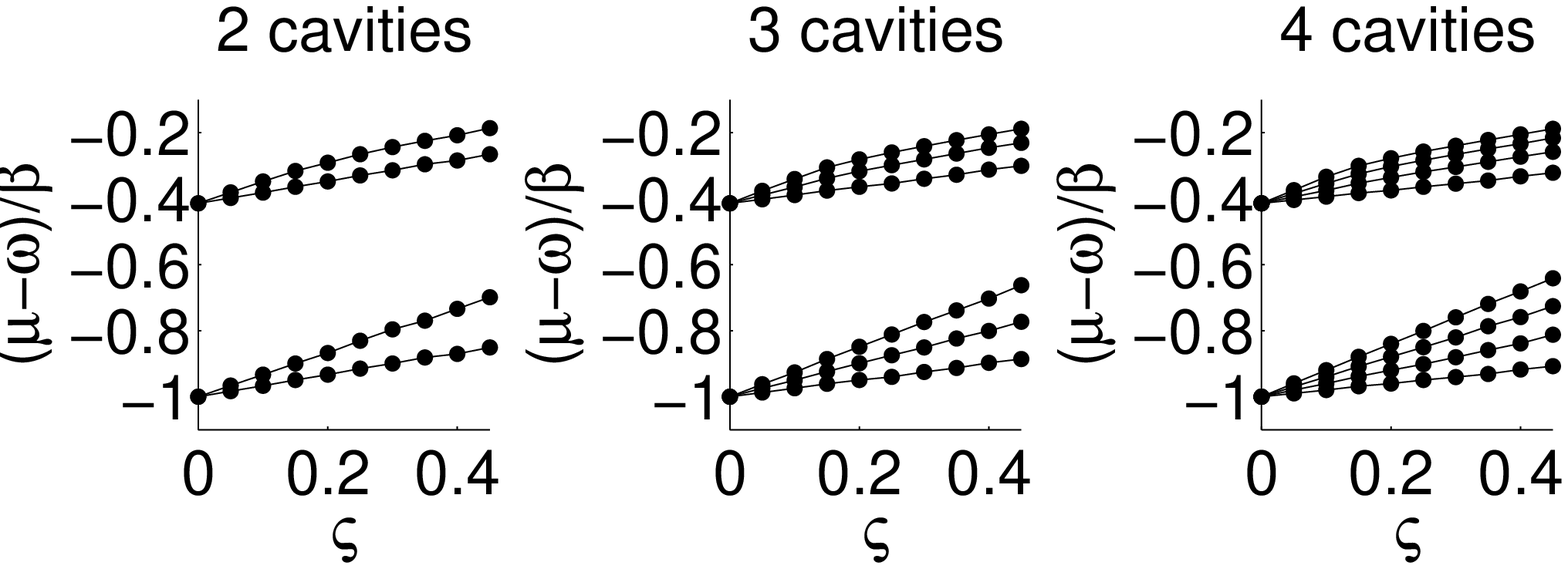}}
  \caption{(a) shows how the boundaries between plateaus change (when
    $\kappa=0$) for 10, 100, and 1000 cavities with increasing
    temperature in natural units.  (b) shows how the boundaries
    between plateaus change (when $\kappa=0$) for two, three, and four
    cavities with increasing disorder, measured in units of standard
    deviation $\varsigma$.  Note that when $\varsigma=0.4$, the
    difference between the top line of the bottom pinch, and the
    bottom line of the top pinch, is $(\mu-\omega)/\beta=$0.434,
    0.361, and 0.320 for two, three and four cavities, respectively.
    This gives some indication of a threshold of tolerance.}
  \label{fig:temperatureanddisorders}
\end{figure}}

\newcommand{\placetstar}{\begin{figure}[t!]
  \centering
  \includegraphics[width=\thecolwidth]{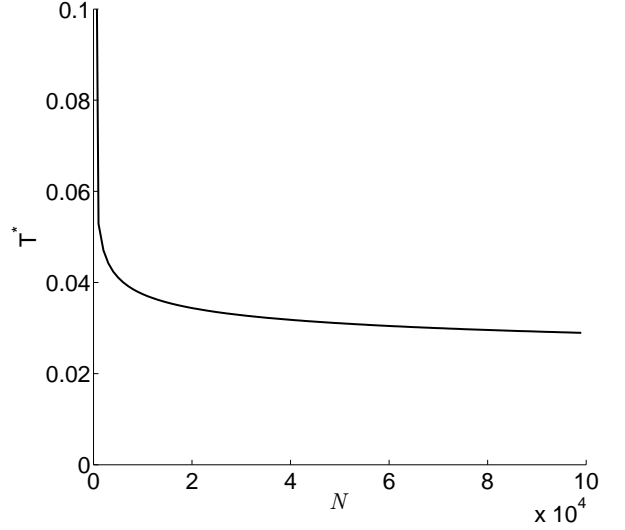}
  \caption{How $T^*$, the temperature at which the top boundary of the
    first group in Fig.~\ref{fig:temperatureanddisorders}(a) meets the
    boundary of the second group, changes as a function of the number
    of cavities.}
  \label{fig:tstar}
\end{figure}}

\title{Quantum phase transitions in photonic cavities with two-level
  systems}

\author{M.I. Makin}
\affiliation{Centre for Quantum Computer Technology, School of
  Physics, The University of Melbourne, Victoria 3010, Australia}

\author{Jared H. Cole}
\affiliation{Centre for Quantum Computer Technology, School of
  Physics, The University of Melbourne, Victoria 3010, Australia}

\author{Charles Tahan}
\affiliation{Cavendish Laboratory, University of Cambridge, JJ Thomson Ave, Cambridge CB3 OHE, United Kingdom}

\author{Lloyd C.L. Hollenberg}
\affiliation{Centre for Quantum Computer Technology, School of
  Physics, The University of Melbourne, Victoria 3010, Australia}

\author{Andrew D. Greentree}
\affiliation{Centre for Quantum Computer Technology, School of
  Physics, The University of Melbourne, Victoria 3010, Australia}

\begin{abstract}

  Systems of coupled photonic cavities have been predicted to exhibit
  quantum phase transitions by analogy with the Hubbard model.  To
  this end, we have studied topologies of a few (up to six) photonic
  cavities each containing a single two-level system.  Quantum phase
  space diagrams are produced for these systems, and compared to
  mean-field results.  We also consider finite effective temperature,
  and compare this to the notion of disorder.  We find the extent of
  the Mott lobes shrink analogously to the conventional Bose-Hubbard
  model.

\end{abstract}

\maketitle

\newcommand{\Ha}{\mathcal{H}}
\newcommand{\Hl}{\Ha^{(l)}}

\section{Introduction}

There has recently been a convergence of several different fields of
physics: condensed matter, quantum optics and information science.
This convergence has been realized by a staggering increase in the
ability to fabricate and control quantum systems experimentally, and
an ability to attack theoretical problems of increasing complexity.
One aspect of this convergence of fields is the push to realize the
quantum computer. Here we discuss another aspect: that of a quantum
simulator.  In particular, we explore the possibility for a quantum
atomic-optical system (here an interacting lattice of optical cavities
with embedded two-state systems) to undergo a quantum phase transition
by direct analogy with the Hubbard model.

The Hubbard model \cite{ref:hubbard} describes the hopping of
interacting particles around a lattice of allowed positional states. A
quantum phase transition is observed between delocalized particles
(superfluid phase) and localized particles (Mott-insulator phase)
depending on the strength of the hopping term relative to the onsite
interaction.  Numerous facets of the Hubbard model have been
considered including the prediction of glassy phases
\cite{ref:fisher}, Hilbert-space optimization
\cite{ref:greentreelots}, and implementations of topological quantum
computing \cite{ref:freedmannayakshtengel}.  One of the most dramatic
and beautiful examples of the Bose-Hubbard model is the prediction
\cite{ref:jaksch} and recent demonstration \cite{ref:greiner2bloch} of
the quantum phase transition in an ultra-cold atomic gas.  Such
demonstrations are significant for applying canonical solid-state
treatments to the more controllable regime of atom optics, allowing
new predictions to be tested (e.g.~the supersolid phase
\cite{ref:scarola}). Recent work on quantum phase transitions in
photonic band-gap lattices does the same for the photonics-solid-state
boundary \cite{ref:hartmannNP, ref:GTCH,ref:angelakis07,
  ref:pointylobes, ref:hartmann08twocomponent, ref:na2yamamoto}.

At first glance, the possibility for a quantum phase transition in an
optical system seems surprising.  This is due to the fact that photons
do not normally interact with each other with any appreciable
strength, and for this reason most nonlinear optical processes are
confined to the realm of classical optics.  There are, however, many
exceptions to this, but perhaps the most dramatic is the phenomenon of
photon blockade \cite{ref:imamoglu}.  Photon blockade is an example of
a cavity quantum electrodynamical interaction in the strong coupling
limit. An atom is placed in a cavity, and because the energy levels of
the atom-cavity system depend on the number of photons in the cavity,
a photon-number dependent resonance shift is observed
\cite{ref:carmichael}. If the atom-cavity interaction is strong
enough, this shift can be sufficient to prevent more than a
predetermined number of photons to enter the cavity: photon blockade.
This effect has been analyzed for four-state systems
\cite{ref:imamoglu, ref:gheri, ref:greentreeBlockade} and two-state
systems \cite{ref:rebic}.  More recently photon-blockade has been
observed \cite{ref:birnbaum, ref:dayan}, adding substantial impetus to
apply this effect to a range of applications.

Here we consider the properties of a lattice of cavities, each
containing a single, quasi-resonant two-state system, so as to be
effectively treated by the Jaynes-Cummings interaction
\cite{ref:jaynescummings}. We go beyond the earlier idealized
treatment \cite{ref:GTCH} by building systems of increasing size to
predict the results in the few-cavity (up to six) limit, and also
consider the thermodynamic implications of disorder. By directly
connecting these small scale cases (solved by direct diagonalization)
with the thermodynamic limit, our results serve as a guide to coming
proof-of-concept experiments. We are also able to compare our finite
cases with the thermodynamic limit \cite{ref:GTCH, ref:na2yamamoto,
  ref:sheshadri, ref:vanoosten, ref:hartmann08migration, ref:leebulla,
  ref:sengupta}.

In Sec.~\ref{sec:introham} we introduce the system of coupled photonic
blockade cavities that will be investigated for quantum phase
transitions, and the extended Jaynes-Cummings-Hubbard Hamiltonians for
both the exact calculation and for the mean-field approximation. In
Sec.~\ref{sec:results} we present results from the exact
diagonalization techniques and compare with mean-field
solutions. Finally, we consider disorder and the implications for
effective model temperature in Sec.~\ref{sec:disordertemperature}.

\section{Phase transitions in the Jaynes-Cummings-Hubbard model}
\label{sec:introham}

The system under consideration is a lattice of optical cavities, each
containing a single quasi-resonant two-state system.  The canonical
treatment for a single atom-cavity system is the Jaynes-Cummings
model. Photon hopping between cavities (which is effected by leakage
out of the cavities, and into neighboring cavities) allows the direct
comparison to Hubbard systems \cite{ref:sachdev}, and hence we refer
to this as a Jaynes-Cummings-Hubbard (JCH) model.  There are numerous
ways in which to realize such a system depending on the available
experimental configurations and desired topologies; for example,
photonic band-gap structures \cite{ref:GTCH} and coupled-cavity
waveguides \cite{ref:angelakis07, ref:pointylobes}, perhaps realized
in micro-fabricated diamond \cite{ref:oliveroetal, ref:tomalphabet,
  ref:GreentreeJPCM}, arrays of superconducting strip-line cavities
\cite{ref:wallraff, ref:schuster}, or microcavities with individual
cold-atoms connected via optical fiber interconnects
\cite{ref:trupke}, or plasmonics \cite{ref:changetal}.  For
concreteness, we will focus our attention on a photonic bandgap
structure where a two-dimensional array of photonic bandgap cavities
constitutes the underlying lattice and defines the nearest-neighbor
topology, and the two-state system is realized by an implanted
impurity, Fig.~\ref{fig:implementation}.

\placeimplementation

Our emphasis in this work is on systems with a single two-state system
per cavity, but it is important to note that this is not the only
potential system for observing similar quantum phase transitions.
Hartmann \textit{et al.}~have considered four-state systems
\cite{ref:hartmannNP, ref:hartmannPRL}, in keeping with the original
Imamo\v{g}lu proposal \cite{ref:imamoglu}, whereas the case of many
atoms per cavity has been considered by Na \textit{et al.}
\cite{ref:na2yamamoto}.  The approach of Ref.~\cite{ref:na2yamamoto} is
particularly useful for providing a clear path to experiments using
GaAs quantum dots.

To understand the properties of the JCH system, we first review the
properties of the individual atom-cavity (Jaynes-Cummings)
interactions.  The Hamiltonian is
\begin{equation}
\label{eq:JCham}
\Ha^{\rm JC} = \epsilon \sigma^+ \sigma^- + \omega a^{\dagger} a + \beta (\sigma^+ a + \sigma^- a^{\dagger}),
\end{equation}
where $\sigma^+$ and $\sigma^-$ ($a^{\dagger}$ and $a$) correspond to
the atomic (photonic) raising and lowering operators, respectively.
The transition energy of the atomic system is $\epsilon$, the cavity
resonance is $\omega$, and the cavity-mediated atom-photon coupling is
$\beta$.  The difference $\Delta = \omega - \epsilon$ is the detuning.

Let $|g,n\rangle$ ($|e,n\rangle$) ($n \in \mathbb{Z}^*$) represent a
cavity that contains $n$ photons and a single two-level atom in the
ground (excited) state.  The energy eigenvectors of
Eq.~(\ref{eq:JCham}) are given by $|g,0\rangle$ and
\begin{equation}
\begin{split}
 |\pm,n\rangle = \frac{\beta\sqrt{n} |g,n\rangle + [-(\Delta/2)\pm \chi(n)]|e,n-1\rangle}{\sqrt{2\chi^2(n)\mp \chi(n)\Delta}}\\
 \quad \forall n\geq 1,
\end{split}
\end{equation}
with eigenvalues
\begin{equation}
\label{eq:singlecavityeigenvalues}
E_{|g,0\rangle} = 0,\quad E_{|\pm,n\rangle} = n \omega \pm \chi(n) - \Delta/2,
\end{equation}
where we have used the generalized Rabi frequency

\begin{equation}
\label{eq:chi}
\chi(n) = \sqrt{n\beta^2 + \Delta^2/4}\quad \forall n\geq 1
\end{equation}
where $n$ is the total number of excitations.  These eigenstates
correspond to the well known dressed (polaritonic) states, and we call
the basis formed by them the single-cavity dressed basis.  The
eigenspectrum of a single atom-cavity system is shown in
Fig.~\ref{fig:JCEig}.  Because of the atom-photon induced shift of the
energy levels as a function of the number of excitations in the
system, there is an effective photon-photon repulsion
\cite{ref:rebic}.  It is this photon-photon repulsion which plays the
role of the on-site term in the Hubbard model, however it is important
to note that because the repulsion decreases with an increasing number
of particles, the canonical Bose-Hubbard system is \emph{not} realized
in our case, and so although many qualitative similarities are
predicted between the JCH and Bose-Hubbard models, exact equivalence
is not guaranteed.

The non-bosonic nature of the particles in the Hamiltonian of
Eq.~(\ref{eq:JCham}) requires further discussion.  Neither the JCH
system nor the four state system with few atoms per cavity
\cite{ref:hartmannNP} retrieve bosonic commutation relations.  The
limits where we can view the system as being comprised of interacting
bosons are many atoms per cavity (holds for both the Jaynes-Cummings
and four state systems \cite{ref:na2yamamoto, ref:irish,
  ref:hartmannNP}), large detuning \cite{ref:GTCH}, and large
excitation number.  Arguably the most important case is that described
here, namely one atom per cavity with few excitations.  This is
because this regime maximizes the non-linear (photon-photon)
interactions, and is therefore the most experimentally accessible
regime.

Differences between the JCH and Bose Hubbard systems are interesting
topics for investigation, and a study of the particle nature should
prove fruitful, but goes beyond our present work.  We may understand
some of the differences by comparing the onsite repulsion in the Bose
Hubbard and JCH cases.  In the Bose Hubbard system, the interaction
$U$ is a constant, however in the JCH model this can be seen as having
a particle number dependence, i.e.~$U_{\pm}(n) = \chi(n+1)-\chi(n)$.
In the large photon limit, we obtain a non-interacting Bose gas, as
$U_{\pm}(n)\rightarrow0$, and in the large detuning limit,
$U_{\pm}(n)\rightarrow\pm\beta^2/\Delta$, which is a constant bosonic
Hubbard type repulsion \cite{ref:irish, ref:na2yamamoto}.  There is also no
ideal Kerr-type term to generate an exact quartic interaction.
Nonetheless, as has been shown, qualitative similarity between the
phase diagrams of JCH and Bose-Hubbard systems is found, and the
analysis of these phase diagrams is a major topic of this paper.

\placefigone

To generate the JCH Hamiltonian, we add hopping between cavities, and
for a system of $N$ cavities we have
\begin{equation}
\label{eq:mainham}
\Ha = \sum_{i=1}^N \Ha^{\rm JC}_i - \sum_{\langle i,j\rangle} \kappa_{ij}a^{\dagger}_i a_j,
\end{equation}
where individual Jaynes-Cummings Hamiltonians of Eq.~(\ref{eq:JCham})
have identical $\epsilon, \omega$ and $\beta$, (this restriction will
be relaxed later). The intercavity hopping occurs with frequency
$\kappa_{ij}=\kappa$ for nearest neighbors, and $\kappa=0$ otherwise,
it is this term which defines the topology of the network.  Photon
transmission through a one-dimensional chain in a similar structure
has also been considered \cite{ref:CPSun}.

To divine the properties of the phase transition seen in the
thermodynamic limit, we introduce the operator that measures the total
number of excitations of the system $\hat{L} = \sum_{i=1}^N
\hat{L}_i$, where $\hat{L}_i = \sigma^+_i\sigma^-_i + a^{\dagger}_i
a_i$ is the number operator of atomic and photonic excitations of the
$i$th cavity.  One can include a term $-\mu\hat{L}$ in the
Hamiltonian.  We show below, through arguments of statistical
mechanics, that $\mu$ represents the chemical potential.  Section
\ref{sec:disordertemperature} will continue into a discussion on
effective model temperature and disorder. Let us include this chemical
potential term in the Hamiltonian directly as follows: 

\begin{equation}
\label{eq:Ha'}
  \Ha' = \Ha - \mu \hat{L}.
\end{equation}
We assume that the entire $N$ cavity system with $l$ \emph{total}
excitations exists in the ground energy eigenstate $|\psi_{\rm
  g}\rangle$, so that $\Ha' |\psi_{\rm g}\rangle = E_{\rm g}
|\psi_{\rm g}\rangle$ and $\hat{L}|\psi_{\rm g}\rangle = l |\psi_{\rm
  g}\rangle$ (i.e.~these two operators commute).

To show that $\mu$ has the general form of a chemical potential, we
begin with the usual definition of free energy $F=E-TS$, where $E$ is
the energy of the system $\Ha$ (before chemical potential has been
included), $T$ is temperature and $S$ is entropy. Assuming that $T=0$
and using the definition of chemical potential as the derivative of
the free energy with respect to number of excitations
\begin{equation}
\label{eq:mudef}
\mu \equiv \left( \frac{\partial F}{\partial l} \right)_{T,V} = \left( \frac{\partial E}{\partial l} \right)_{T,V},
\end{equation}
where $l$ is used, as excitations act like particles in this system.
We use the Hellmann-Feynman theorem \cite{ref:hellmann,ref:feynman}
to calculate the derivative of the energy with respect to number of
excitations
\begin{equation}
\begin{split}
\frac{\partial E}{\partial l}& = \langle\psi_{\rm g}| \frac{\partial \Ha}{\partial l}|\psi_{\rm g}\rangle, \\
&= \langle \psi_{\rm g}| \frac{\partial }{\partial l}(\Ha'+\mu l) |\psi_{\rm g}\rangle, \\
&= \mu.
\end{split}
\end{equation}
Hence the $\mu$ of Eq.~(\ref{eq:Ha'}) represents chemical potential as
required.


The $N$ cavity bare basis consists of state vectors of the form
$|s_1,n_1\rangle \otimes |s_2,n_2\rangle \otimes \cdots \otimes
|s_N,n_N\rangle$, $s_i \in \{g,e\}$, $n_i \in \{0,1,\ldots\}$.  In
principle, this basis is infinite in extent, because the number of
photonic excitations per cavity is unbounded.  By ordering the bare
basis by the total number of excitations (either photonic or
atomic) across all cavities, one may express the Hamiltonian of
Eq.~(\ref{eq:mainham}) in block diagonal form $\Ha={\rm
  diag}[\Ha^{(0)},\Ha^{(1)},\Ha^{(2)},\ldots]$, where $\Ha^{(l)}$ is
the matrix corresponding to $l$ excitations.  The size of each block
is determined by the number of ways in which the excitations can be
shared between the atomic and photonic degrees of freedom.  We denote
the number of states (equal to the size of the matrix
$\mathcal{H}^{(l)}$) as $s$, where

\begin{equation}
\label{eq:s}
s = \sum_{i=0}^{\min(l,N)} {N \choose i} S_{l-i}^N.
\end{equation}
The above summation has two terms. $N\choose i$ is the total number of
atomic excitations across the lattice (note that on each site the
number of atomic excitations can only be zero or one). $S_i^N$
represents the number of photonic excitations, and is the number of
ways to share the $l-i$ photons between the $N$ cavities, (e.g.
$S_2^3={\rm length}[(2,0,0)$, $(0,2,0)$, $(0,0,2)$, $(1,1,0)$,
$(1,0,1)$, $(0,1,1)]=6$).

To gain insight into the problem, we explicitly show $\Ha^{(0)}$ and
$\Ha^{(1)}$ for a two-cavity system in the bare basis as follows:

\begin{equation}
\begin{split}
  &\Ha^{(0)} = \left( \begin{array}{c}0\end{array}\right)\\
  &\Ha^{(1)} = \left( \begin{array}{cccc}
      \omega & \beta & -\kappa & 0\\
      \beta & \epsilon & 0 & 0\\
      -\kappa &0 & \omega &  \beta\\
      0 & 0 & \beta & \epsilon\\
\end{array}\right),\\
\end{split}
\end{equation}
where the (two-cavity) basis for $\Ha^{(0)}$ is $\{|g,0,g,0\rangle\}$
and for $\Ha^{(1)}$ is $\{|g,1,g,0\rangle, |e,0,g,0\rangle,
|g,0,g,1\rangle, |g,0,e,0\rangle\}$.

To connect the finite case with the thermodynamic limit, we examine
the phases of the $N$ cavity system $\Ha'$ [Eq.~(\ref{eq:Ha'})] as a
function of $\omega$, $\kappa$, and $\Delta$.  In particular, we are
concerned with the expectation value of the total number of
excitations of the system $\langle\hat{L}\rangle$.  Note that
$\hat{L}$ is diagonal when represented in either the dressed or bare
basis (but has different values in each).  From \cite{ref:witte}, by
subtracting $\mu \hat{L}$ from our Hamiltonian Eq.~(\ref{eq:mainham}),
we determine that $\langle \hat{L}\rangle =-\partial E_{\rm
  g}/\partial \mu$, where $E_{\rm g}$ is the ground state energy of
the \emph{extended} Hamiltonian $\Ha'$.  Some preliminary analytics
can simplify the calculation of $\langle\hat{L}\rangle$ considerably;
we show this now.

We begin by noting that the part of $\hat{L}$ corresponding to exactly
$l$ excitations, represented as $\hat{L}^{(l)}$, has the very simple
form

\begin{equation}
\hat{L}^{(l)} = l I.
\end{equation}
Again employing the Hellmann-Feynman theorem,

\begin{equation}
\frac{\partial E_{\rm g}^{(l)}}{\partial \mu} = \langle \psi_{\rm g} |
\frac{\partial}{\partial \mu} (\Ha^{(l)} - \mu l I) |\psi_{\rm
g}\rangle,
\end{equation}
where $E_{\rm g}^{(l)}$ is the ground state energy, and $|\psi_{\rm
  g}\rangle$ is the corresponding eigenstate, of
$\Ha^{(l)}-\mu\hat{L}^{(l)}$.  This reduces to

\begin{equation}
\frac{\partial E_{\rm g}^{(l)}}{\partial \mu} = -l.
\end{equation}
So, if

\begin{equation}
\begin{split}
M = \{&\min[{\rm eigenvalues}(\Ha^{(0)} - \mu\hat{L}^{(0)})],\\
&\min[{\rm eigenvalues}(\Ha^{(1)} - \mu\hat{L}^{(1)})],\ldots\},
\end{split}
\end{equation}
and

\begin{equation}
f:\{0,1,\ldots\}\rightarrow M,
\end{equation}
then

\begin{equation}
\label{eq:finv}
\langle \hat{L}\rangle = f^{-1} [\min(M)].
\end{equation}
In short, to find $\langle \hat{L}\rangle$, one simply needs to locate
which block the minimum eigenvalue of $(\Ha - \mu \hat{L})$
corresponds to.  Obviously, $\langle \hat{L}\rangle$ can only have
non-negative integer values.  This is illustrated further in
Fig.~\ref{fig:energies}.  In this figure, the smallest eigenvalue of
$\mathcal{H}^{(0)}, \mathcal{H}^{(1)}, \ldots$ is plotted as a
function of $(\mu-\omega)/\beta$ for $\kappa/\beta=0$.  For each
value, $\langle\hat{L}\rangle$ is given by the negative slope of the
smallest eigenvalue at that point.

\placeenergies

We now introduce the mean-field Hamiltonian.  The mean-field
approximation focuses attention on one particular cavity, and assumes
that its $z$ nearest neighbors (that is, the coordination number is
$z$) all behave like it.  To invoke the mean-field, we use the
decoupling approximation $a^{\dagger}_i a_j = \langle
a^{\dagger}_i\rangle a_j + \langle a_j\rangle a^{\dagger}_i - \langle
a^{\dagger}_i \rangle \langle a_j\rangle$, and introduce the
superfluid order parameter $\psi=\langle a_i\rangle = \langle
a^{\dagger}_i \rangle$ (which we assume real), so that the Hamiltonian
of Eqs.~(\ref{eq:mainham}) and (\ref{eq:Ha'}) becomes

\begin{equation}
\label{eq:meanfield}
  \Ha^{\rm MF} = \Ha^{\rm JC} - z \kappa \psi (a^{\dagger} + a ) +  z \kappa \psi^2 - \mu\hat{L}.
\end{equation}
The basis uses just one cavity, but the system (approximately)
describes an infinite number.  Note that the number of nearest
neighbors $z$ effectively ``renormalizes'' the mean-field coupling,
i.e.,~$\kappa \rightarrow z \kappa$.

We have also considered using a larger unit cell - for example, using
two cavities, each with $z$ nearest neighbors.  We found that while
this is more difficult to calculate (finding eigenvalues of larger
matrices) this exactly replicates the results of the original
mean-field - which is not surprising.  However, this technique could
be used to include disorder in the infinite cavity limit.  Note that
while Eq.~(\ref{eq:finv}) informs that the \emph{total} number of
excitations is integer, the mean-field Hamiltonian of
Eq.~(\ref{eq:meanfield}) informs of the number of excitations \emph{per
  cavity}.  Accordingly, we will find equivalence when the number of
excitations is a multiple of the number of cavities.

\section{Results}
\label{sec:results}

In this section we analyze the quantum phase diagram of the
Hamiltonian of Eq.~(\ref{eq:mainham}) for various topologies.  Our
analysis is based on exact diagonalization for up to six cavities.
Topology is implemented through the $\kappa_{ij}$ terms of
Eq.~(\ref{eq:mainham}).  We then compare these topologies with the
mean-field approximation.

We begin by considering the quantum phase diagrams of the exact
systems.  We display the phase diagrams corresponding to two, three,
four, and five cavities arranged in one dimension with periodic
boundary conditions in Fig.~\ref{fig:exact}.  Each color corresponds
to a different plateau, a constant state in excitation space - these
are Mott insulating phases.  It is worth pointing out that in our
discrete model, no superfluid phase exists.  However, for
significantly large $\kappa$, the plateaus get closer together,
approximating the superfluid phase diagram, as in the mean-field case.

\placeexact

In total, eleven topologies were examined.  These are listed in the
first column of Table \ref{tab:states}. The topologies of a square,
triangle and six cavities with $z=3$ could be considered special, as
they can represent infinite square, triangular and hexagonal lattices
respectively.  However, no significant differences (with respect to
matching of phase diagrams to mean-field) are found between these
topologies and the rest.

For all topologies, a ``pinch'' effect is noted as
$\kappa\rightarrow0$ between $\langle\hat{L}\rangle=N$ and
$\langle\hat{L}\rangle=2N$, between $\langle\hat{L}\rangle=2N$ and
$\langle\hat{L}\rangle=3N$, etc.  That is, all fractional occupations
(plateaus corresponding to heights that are \emph{not} integer
multiples of the number of cavities $N$) disappear as
$\kappa\rightarrow0$, this compares nicely with the mean-field
solution.  The point at which this pinching occurs is called the
critical chemical potential $\mu_c$. We find

\begin{equation}
\label{eq:criticalchempot}
\mu_c(n) = \omega+\chi(n) - \chi(n+1),
\end{equation}
where $\chi(n)$ is defined in Eq.~(\ref{eq:chi}), and
$\chi(0)=-\Delta/2$.  This is independent of the number or arrangement
of cavities, and independent of whether or not the mean-field
approximation is used, as expected from the $\kappa\rightarrow0$
limit.

Although in general, one cannot analytically determine the positions
of all the boundaries, we find that for all topologies, the first
boundary (between $\langle\hat{L}\rangle=0$ and
$\langle\hat{L}\rangle=1$) is described by the analytic equation

\begin{equation}
\label{eq:lowestboundary}
\frac{\mu-\omega}{\beta} = -\frac{1}{2}\left[\frac{\Delta + z\kappa}{\beta} + \sqrt{\left(\frac{\Delta-z\kappa}{\beta}\right)^2+4}\right],
\end{equation}
where $z$ is the number of nearest neighbors (the third column in
Table \ref{tab:states}).  Equation (\ref{eq:lowestboundary}) was
determined by equating the smallest eigenvalue of $\mathcal{H}^{(1)}$
with zero (the only eigenvalue of $\mathcal{H}^{(0)}$).  This has been
compared numerically for $z=1,2,3,4,6$ and is in excellent agreement.
One cannot expect generic $z$ analytic boundaries between
$\langle\hat{L}\rangle=1$ and $\langle\hat{L}\rangle=2$ or higher to
exist, indeed none were determined.  This is because they are truly in
the realm of many-body physics (unlike the lower boundary).  A higher
boundary would be the solution of a commensurately higher order
polynomial.  In the simplest case, $N=2$ and $l=2$; from
Eq.~(\ref{eq:s}), $s=8$, so an eighth order polynomial must be solved
to determine the boundaries.

For each plateau, the ground eigenstate can be calculated (hence below
we use ``ground eigenstates'' to refer to the different ground states
for the different plateaus).  These ground eigenstates are represented
in the dressed state basis, even though many of the calculations above
were done in the bare basis.

To easily represent this information, it is useful to define two
operators: the translation operator $\hat{T}$ (which shifts states to
the right and moves the last state back to the beginning) and the
permutation operator $\hat{P}_m$ (which is the sum over $\hat{T}^i$
applied to the state $m$ times)

\begin{equation}
\label{eq:superstates}
\begin{split}
\hat{T}& \left( |s_1,n_1\rangle \otimes \ldots \otimes |s_{N-1},n_{N-1}\rangle \otimes |s_N,n_N\rangle\right) \\
&= |s_N,n_N\rangle \otimes |s_1,n_1\rangle \otimes  \ldots \otimes |s_{N-1},n_{N-1}\rangle,
\end{split}
\end{equation}
\begin{equation}
\begin{split}
\hat{P}_m& \left( |s_1,n_1\rangle \otimes \ldots  \otimes |s_N,n_N\rangle\right)\\
&=  \sum_{i=0}^{m-1} (\hat{T})^i \left(|s_1,n_1\rangle \otimes \ldots  \otimes |s_N,n_N\rangle\right).
\end{split}
\end{equation}

The ground eigenstates for all eleven topologies, up to
$\langle\hat{L}\rangle=N$, i.e.,~total number of excitations equal to
the number of cavities, are displayed in Table \ref{tab:states}. See
Fig.~\ref{fig:2cavsExplain} for a diagrammatical example of how, for
two cavities, the information from the table matches a phase diagram.

\placetwocavsExplain

Consider bands labeled by $m$, where

\begin{equation}
\label{eq:band}
m N \leq \langle\hat{L}\rangle \leq (m+1) N -1\quad \forall m=0,1,\ldots.
\end{equation}
One finds that the physics of each band has some striking
similarities, hence the introduction of this parameter.

In Table \ref{tab:states} only results for $m=0$ are displayed.  While
higher bands include many more possible states (e.g.,~$|-,1\rangle
\otimes|+,1\rangle \otimes|-,2\rangle$), we find higher bands have a
surprisingly simple structure. To obtain the general states for some
other band $m$, one needs to simply replace $|-,1\rangle$ by
$|-,m+1\rangle$, and $|g,0\rangle$ by $|-,m\rangle$, in every
instance.

\newcommand{\firstcolwidth}{0.7cm}

\begin{table*}[htbp]
  \centering
\begin{sideways}
  \begin{tabular}{|c|c|c|c|c|}
    \hline
    Topology & $N$ & $z$ & $\langle\hat{L}\rangle$ & Associated ground eigenstate (dressed basis)\\    \hline
    \hline
    arbitrary &$n$ & N/A &  0 & $|g,0\rangle^{\otimes n}$ \\
    \hline
    arbitrary& $n$ & N/A  &1 & $\frac{1}{\sqrt{n}}\hat{P}_n(|g,0\rangle^{\otimes(n-1)}\otimes|-,1\rangle)$ \\
    \hline
    arbitrary& $n$ & N/A  & $n-1$ & $\frac{1}{\sqrt{n}}\hat{P}_n(|g,0\rangle\otimes|-,1\rangle^{\otimes(n-1)})$ \\
    \hline
    arbitrary  &$n$ & N/A & $n$ & $|-,1\rangle^{\otimes n}$ \\
    \hline
\includegraphics[scale=0.07]{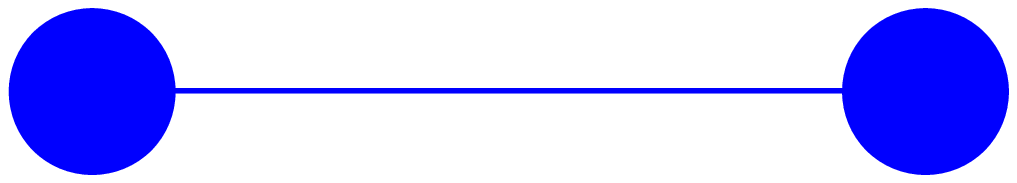}& 2 & 1 & 1 & $\frac{1}{\sqrt{2}}\hat{P}_2(|g,0\rangle\otimes|-,1\rangle)$ \\
\hline
\includegraphics[scale=0.05]{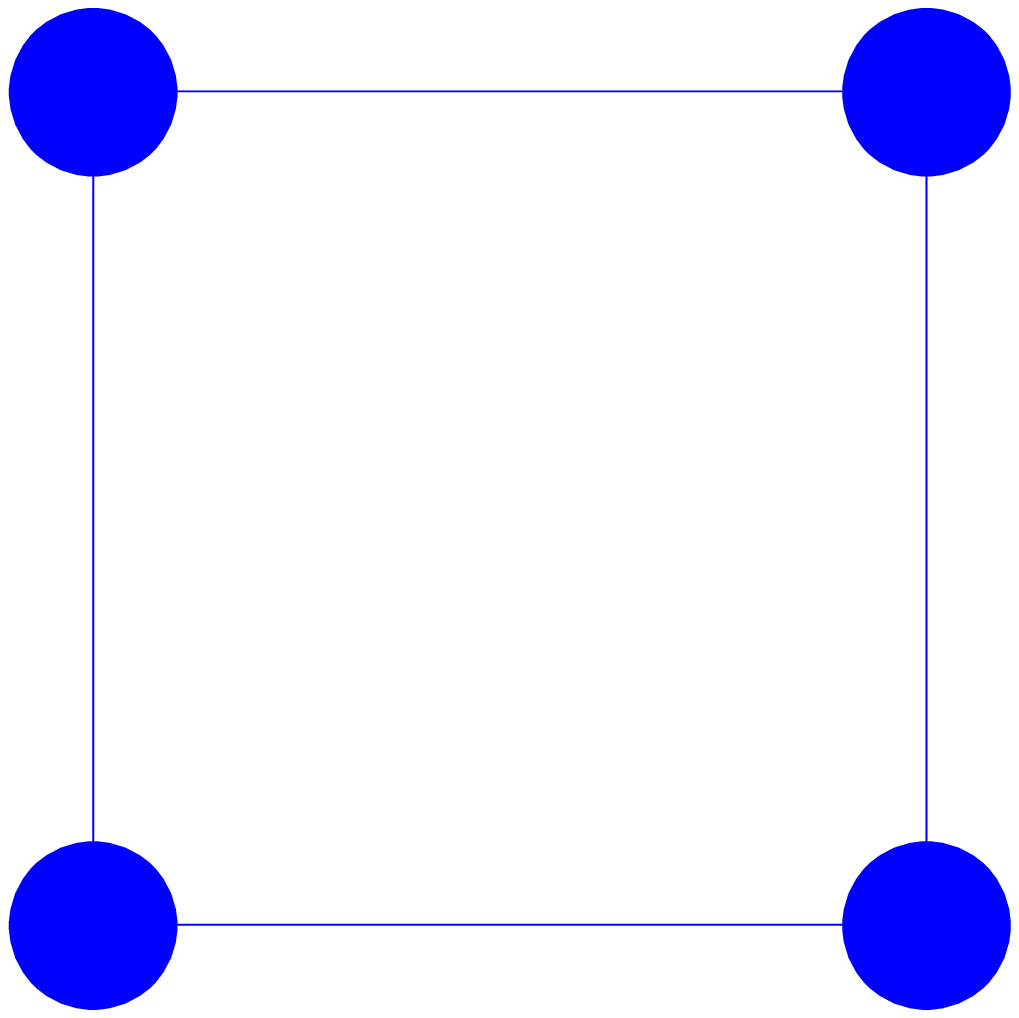}  & 4 & 2 & 2 & $\frac{1}{\sqrt{8}}\hat{P}_4(|g,0\rangle^{\otimes2}\otimes|-,1\rangle^{\otimes2})+\frac{1}{2} \hat{P}_2[(|g,0\rangle\otimes|-,1\rangle)^{\otimes2}]$  \\
\hline
\includegraphics[scale=0.08]{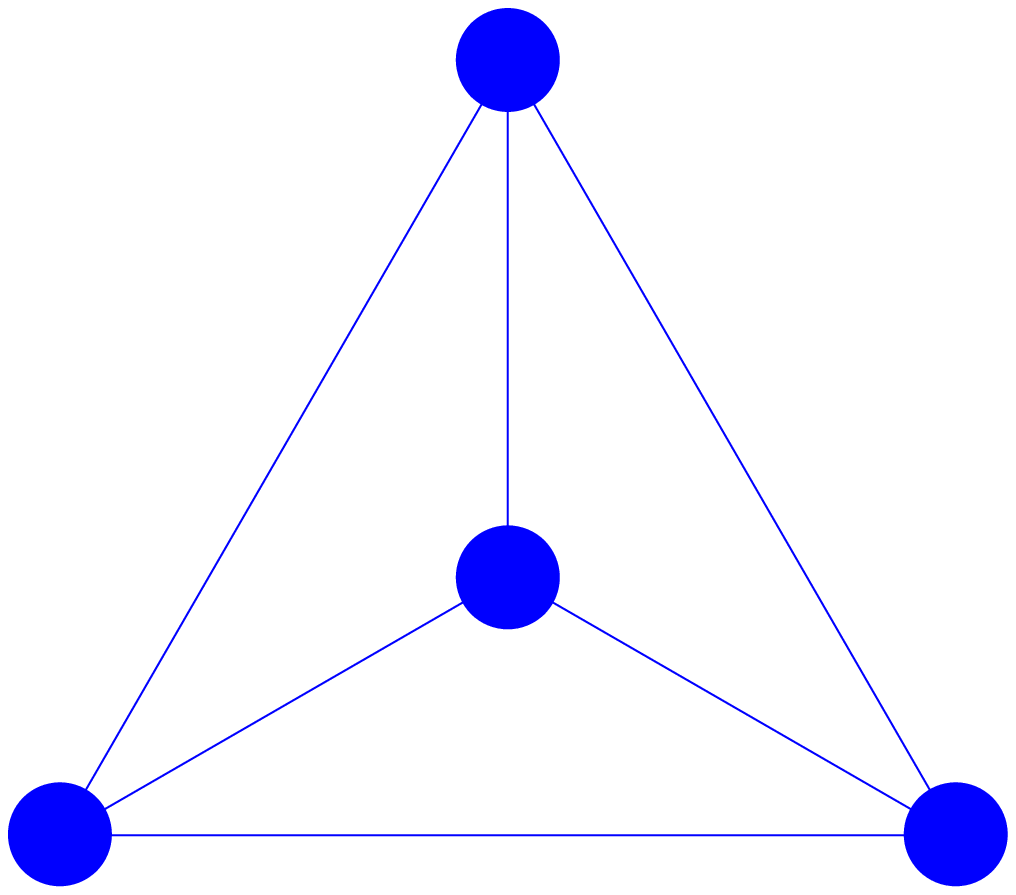}  & 4 & 3 & 2 & $\frac{1}{2}\{\hat{P}_4(|g,0\rangle^{\otimes2}\otimes|-,1\rangle^{\otimes2})+\hat{P}_2[(|g,0\rangle\otimes|-,1\rangle)^{\otimes2}]\}$ \\
\hline
\multirow{2}{\firstcolwidth}{\centering \includegraphics[scale=0.08]{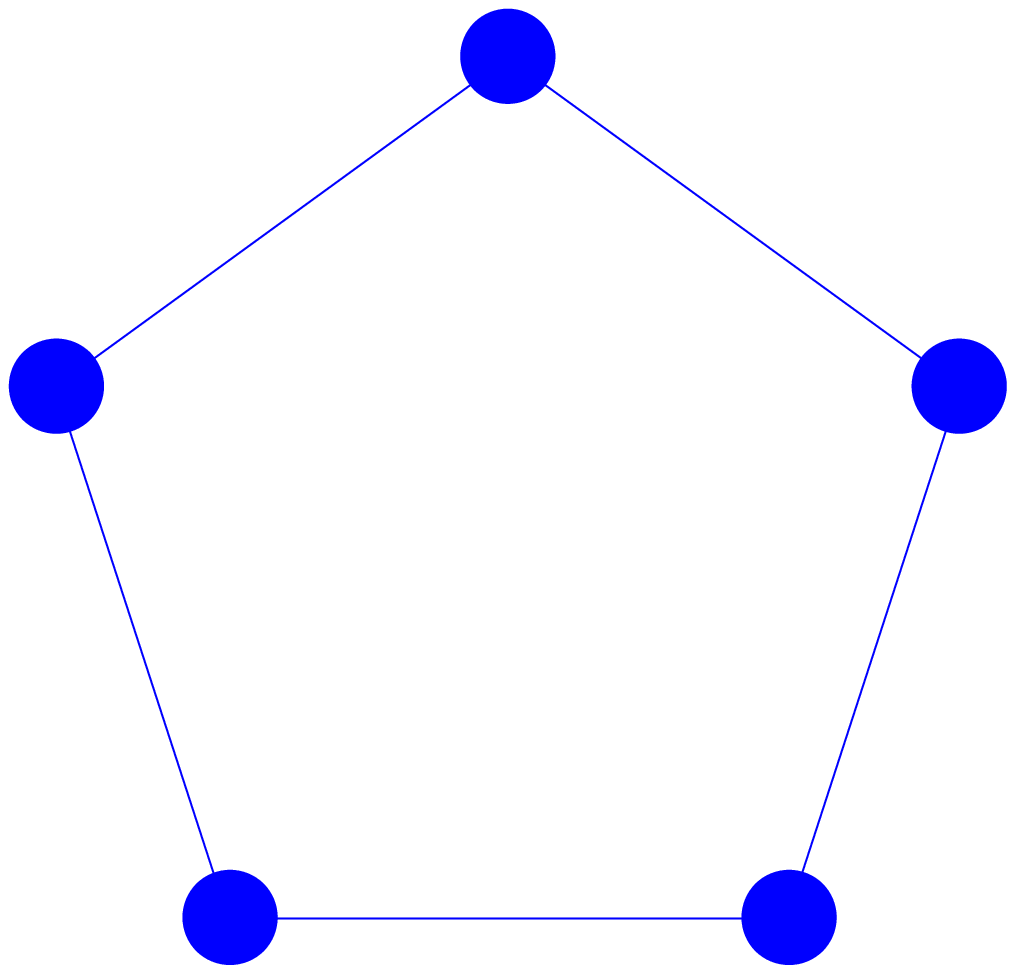}} & \multirow{2}{1cm}{\centering 5} & \multirow{2}{1cm}{\centering 2} & 2 & $0.235\; \hat{P}_5 (|g,0\rangle^{\otimes3}\otimes|-,1\rangle^{\otimes2}) + 0.380\; \hat{P}_5 (|g,0\rangle^{\otimes2}\otimes|-,1\rangle \otimes|g,0\rangle \otimes |-,1\rangle)$ \\
 & & & 3 & $0.235\; \hat{P}_5 (|g,0\rangle^{\otimes2}\otimes|-,1\rangle^{\otimes3}) + 0.380 \;\hat{P}_5(|g,0\rangle\otimes|-,1\rangle\otimes|g,0\rangle\otimes|-,1\rangle^{\otimes2})$ \\
\hline
\multirow{2}{\firstcolwidth}{\centering \includegraphics[scale=0.08]{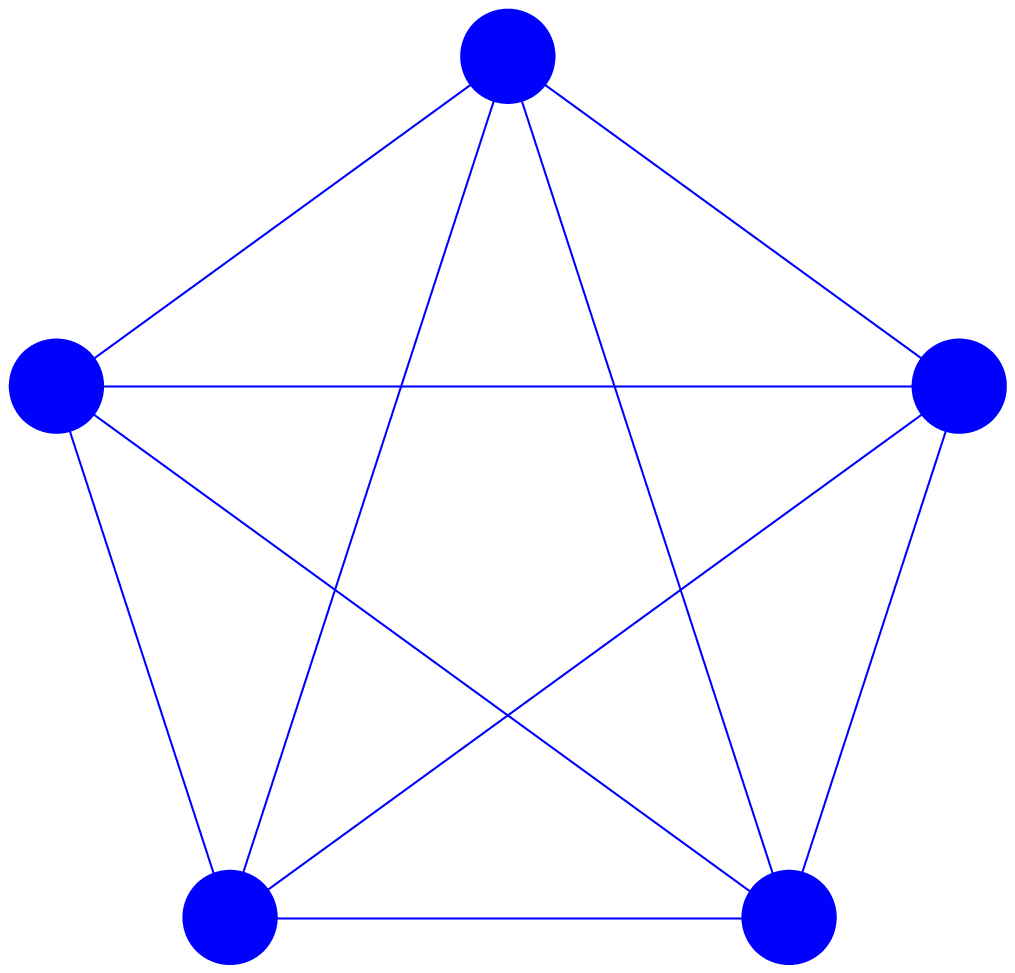}} & \multirow{2}{1cm}{\centering 5} & \multirow{2}{1cm}{\centering 4} & 2 & $\frac{1}{\sqrt{5}}[ \hat{P}_5 (|g,0\rangle^{\otimes3}\otimes|-,1\rangle^{\otimes2}) +\hat{P}_5(|g,0\rangle^{\otimes2}\otimes|-,1\rangle\otimes|g,0\rangle\otimes|-,1\rangle)] $ \\
 & & & 3 & $\frac{1}{\sqrt{5}} [\hat{P}_5(|g,0\rangle^{\otimes2}\otimes|-,1\rangle^{\otimes3}) + \hat{P}_5 (|g,0\rangle\otimes|-,1\rangle\otimes|g,0\rangle\otimes|-,1\rangle^{\otimes2})]$\\
\hline
\multirow{3}{\firstcolwidth}{\centering \includegraphics[scale=0.07]{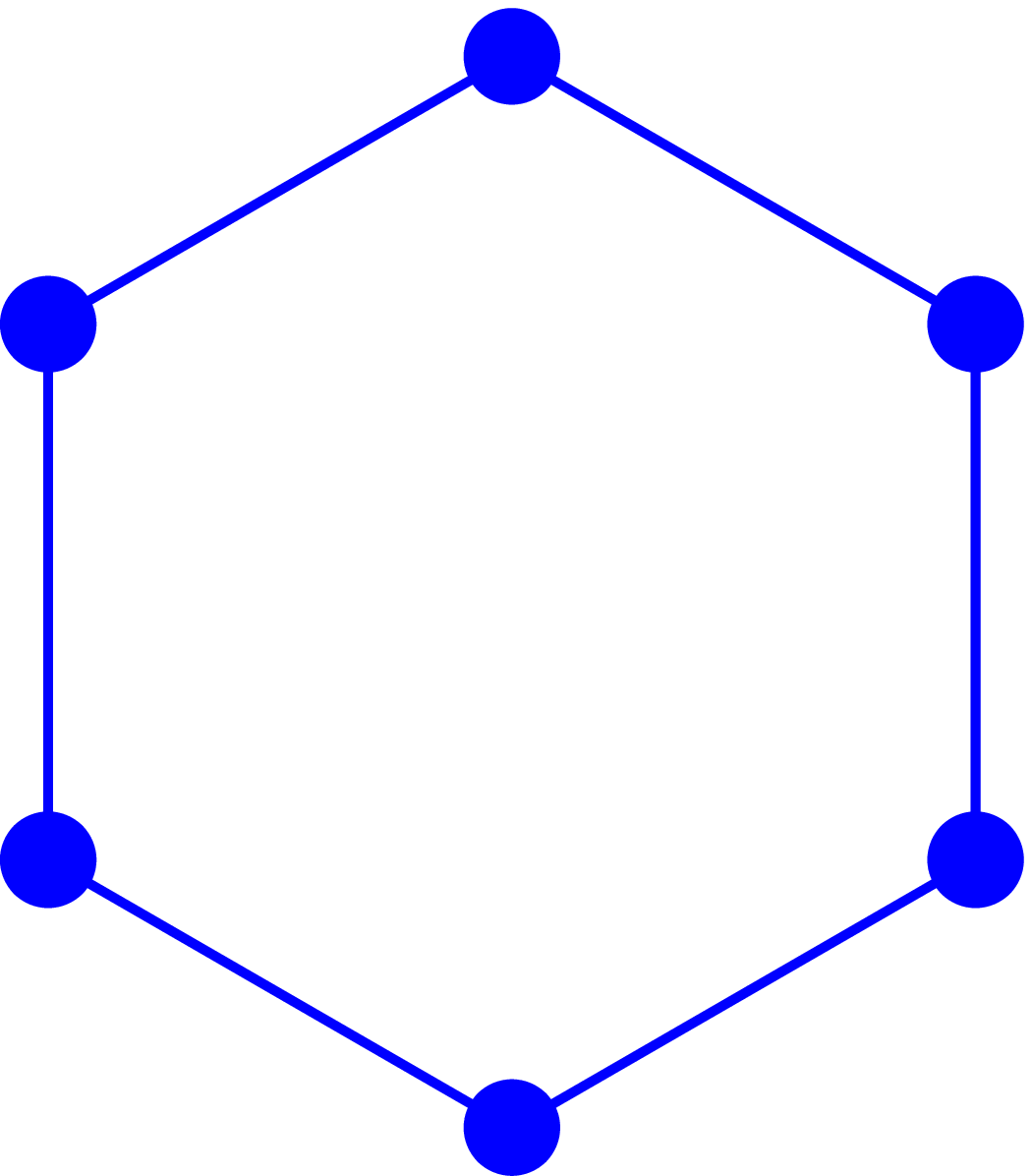}} &\multirow{3}{1cm}{\centering 6} & \multirow{3}{1cm}{\centering 2}& 2 & $\frac{1}{6} \hat{P}_6 (|g,0\rangle^{\otimes4}\otimes|-,1\rangle^{\otimes2}) + \frac{1}{\sqrt{12}} \hat{P}_6 (|g,0\rangle^{\otimes3} \otimes|-,1\rangle\otimes|g,0\rangle\otimes|-,1\rangle) + \frac{1}{3} \hat{P}_3 [(|g,0\rangle^{\otimes2}\otimes|-,1\rangle)^{\otimes2}]$ \\
 &  &  & 3 & $\frac{1}{\sqrt{72}} \hat{P}_6 (|g,0\rangle^{\otimes3}\otimes|-,1\rangle^{\otimes3}) + \frac{1}{\sqrt{18}}  \hat{P}_6 (|g,0\rangle^{\otimes2}\otimes|-,1\rangle\otimes|g,0\rangle\otimes|-,1\rangle^{\otimes2} + |g,0\rangle^{\otimes2}\otimes|-,1\rangle^{\otimes2}\otimes|g,0\rangle\otimes|-,1\rangle) + \frac{1}{\sqrt{8}} \hat{P}_2 [(|g,0\rangle\otimes|-,1\rangle)^{\otimes3}]$ \\
 & & & 4 & $\frac{1}{6} \hat{P}_6 (|g,0\rangle^{\otimes2}\otimes|-,1\rangle^{\otimes4}) + \frac{1}{\sqrt{12}} \hat{P}_6 (|g,0\rangle\otimes|-,1\rangle\otimes|g,0\rangle\otimes|-,1\rangle^{\otimes3}) + \frac{1}{3} \hat{P}_3 [(|g,0\rangle\otimes|-,1\rangle^{\otimes2})^{\otimes2}]$ \\
\hline
\multirow{3}{\firstcolwidth}{\centering \includegraphics[scale=0.07]{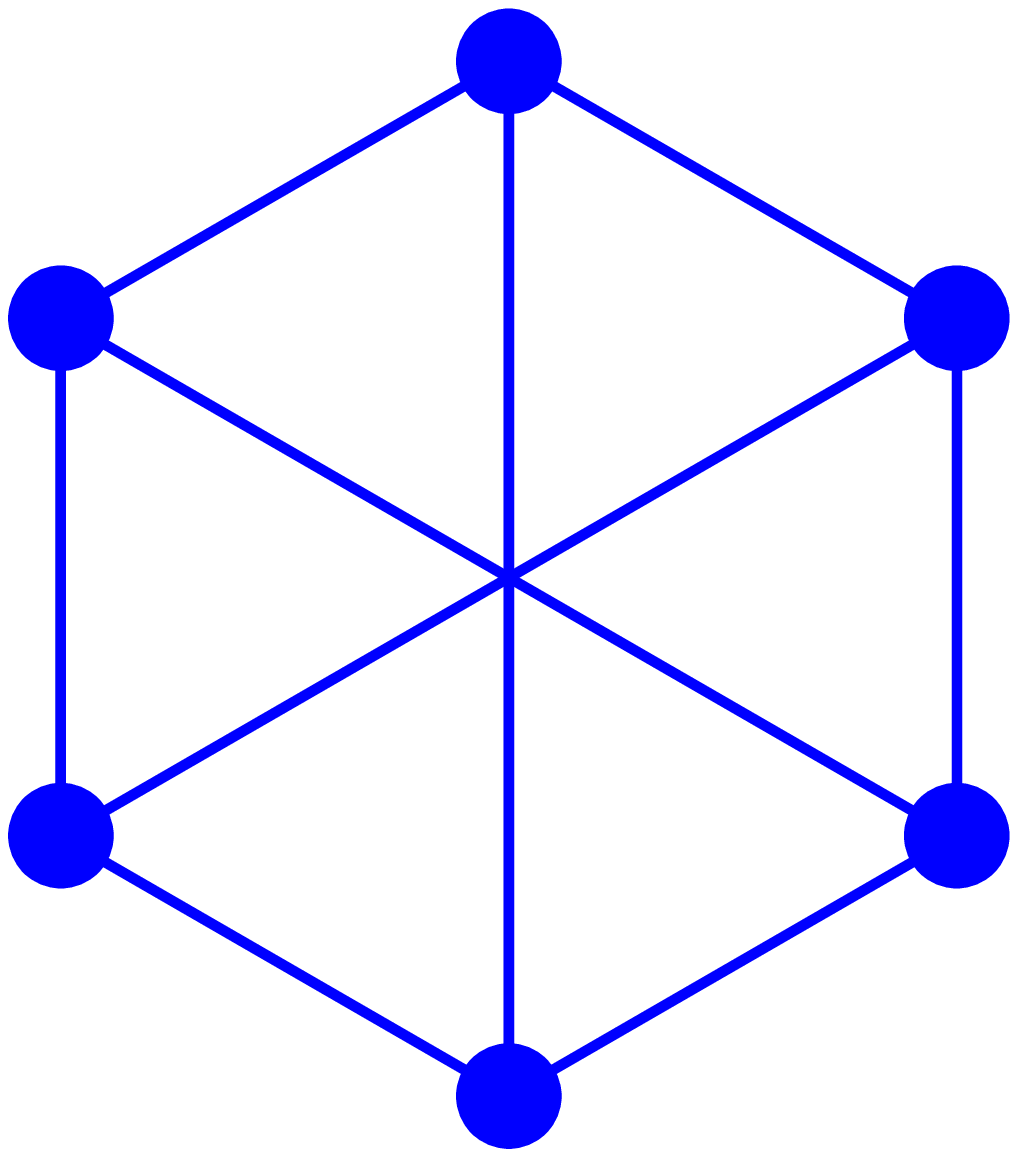}} &\multirow{3}{1cm}{\centering 6} &\multirow{3}{1cm}{\centering 3} & 2 & $\frac{1}{\sqrt{18}} \{\hat{P}_6 (|g,0\rangle^{\otimes4}\otimes|-,1\rangle^{\otimes2}) + \hat{P}_3 [(|g,0\rangle^{\otimes2}\otimes|-,1\rangle)^{\otimes2}\} + \frac{1}{\sqrt{12}} \hat{P}_6 (|g,0\rangle^{\otimes3}\otimes|-,1\rangle\otimes|g,0\rangle\otimes|-,1\rangle)$ \\
 & &  & 3 & $0.208 \hat{P}_6 (|g,0\rangle^{\otimes3}\otimes|-,1\rangle^{\otimes3} + |g,0\rangle^{\otimes2}\otimes|-,1\rangle\otimes|g,0\rangle\otimes|-,1\rangle^{\otimes2} + |g,0\rangle^{\otimes2}\otimes|-,1\rangle^{\otimes2}\otimes|g,0\rangle\otimes|-,1\rangle)  + 0.334 \; \hat{P}_2 [(|g,0\rangle\otimes|-,1\rangle)^{\otimes3}]$\\
 & & & 4 & $\frac{1}{\sqrt{18}} \{\hat{P}_6 (|g,0\rangle^{\otimes2}\otimes|-,1\rangle^{\otimes4}) + \hat{P}_3 [(|g,0\rangle\otimes|-,1\rangle^{\otimes2})^{\otimes2}\} + \frac{1}{\sqrt{12}} \hat{P}_6 (|g,0\rangle\otimes|-,1\rangle\otimes|g,0\rangle\otimes|-,1\rangle^{\otimes3})$\\
\hline
\multirow{3}{\firstcolwidth}{\centering \includegraphics[scale=0.07]{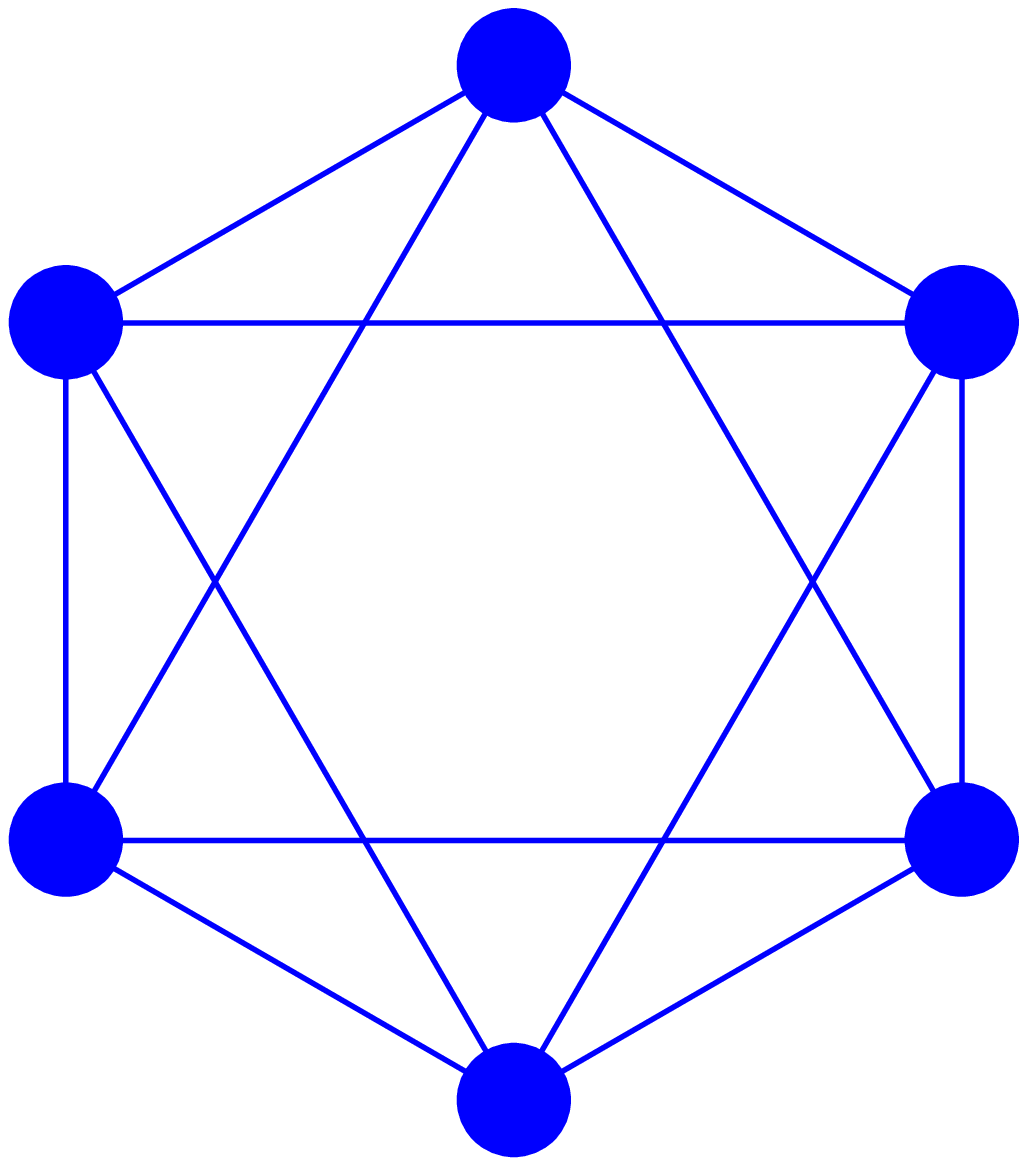}} &\multirow{3}{1cm}{\centering 6} & \multirow{3}{1cm}{\centering 4}& 2 & $0.246 \hat{P}_6 (|g,0\rangle^{\otimes4}\otimes|-,1\rangle^{\otimes2} + |g,0\rangle^{\otimes3}\otimes|-,1\rangle\otimes|g,0\rangle\otimes|-,1\rangle) + 0.304 \; \hat{P}_3 [(|g,0\rangle^{\otimes2}\otimes|-,1\rangle)^{\otimes2}]$\\
 &  &  & 3 & $0.197 \{\hat{P}_6 (|g,0\rangle^{\otimes3}\otimes|-,1\rangle^{\otimes3}) + \hat{P}_2 [(|g,0\rangle\otimes|-,1\rangle)^{\otimes3}]\} + 0.240\;  \hat{P}_6 (|g,0\rangle^{\otimes2}\otimes|-,1\rangle\otimes|g,0\rangle\otimes|-,1\rangle^{\otimes2} + |g,0\rangle^{\otimes2}\otimes|-,1\rangle^{\otimes2}\otimes|g,0\rangle\otimes|-,1\rangle)$\\
 & & & 4 & $0.246 \hat{P}_6 (|g,0\rangle^{\otimes2}\otimes|-,1\rangle^{\otimes4} + |g,0\rangle\otimes|-,1\rangle\otimes|g,0\rangle\otimes|-,1\rangle^{\otimes3}) + 0.304 \;\hat{P}_3 [(|g,0\rangle\otimes|-,1\rangle^{\otimes2})^{\otimes2}]$ \\
\hline
\multirow{3}{\firstcolwidth}{\centering \includegraphics[scale=0.07]{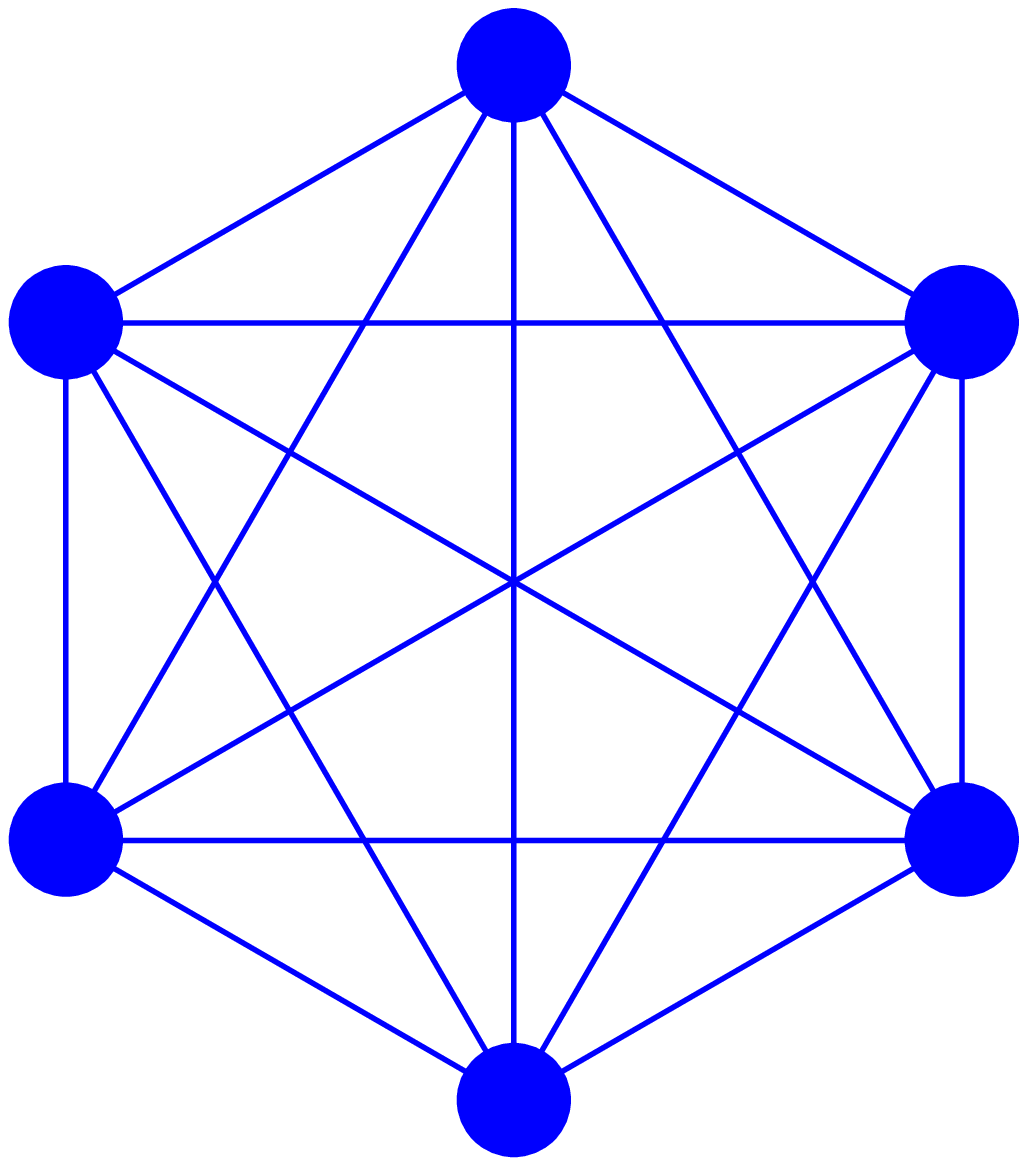}} &\multirow{3}{1cm}{\centering 6} &\multirow{3}{1cm}{\centering 5} & 2 & $\frac{1}{\sqrt{15}} \{\hat{P}_6 (|g,0\rangle^{\otimes4}\otimes|-,1\rangle^{\otimes2} + |g,0\rangle^{\otimes3}\otimes|-,1\rangle\otimes|g,0\rangle\otimes|-,1\rangle) + \hat{P}_3 [(|g,0\rangle^{\otimes2}\otimes|-,1\rangle)^{\otimes2}]\}$\\
 & &  & 3 & $\frac{1}{\sqrt{20}} \{\hat{P}_6 (|g,0\rangle^{\otimes3}\otimes|-,1\rangle^{\otimes3} +  |g,0\rangle^{\otimes2}\otimes|-,1\rangle\otimes|g,0\rangle\otimes|-,1\rangle^{\otimes2} + |g,0\rangle^{\otimes2}\otimes|-,1\rangle^{\otimes2}\otimes|g,0\rangle\otimes|-,1\rangle) + \hat{P}_2 [(|g,0\rangle\otimes|-,1\rangle)^{\otimes3})]\}$\\
 & & & 4 &  $\frac{1}{\sqrt{15}} \{\hat{P}_6 (|g,0\rangle^{\otimes2}\otimes|-,1\rangle^{\otimes4}) + \hat{P}_3[(|g,0\rangle\otimes|-,1\rangle^{\otimes2})^{\otimes2}]+\hat{P}_6 (|g,0\rangle\otimes|-,1\rangle\otimes|g,0\rangle\otimes|-,1\rangle^{\otimes3})\}$\\
\hline
  \end{tabular}
\end{sideways}
\caption{Summary of information about the number of cavities $N$, and nearest neighbors $z$, excitation state $\langle\hat{L}\rangle$ and  associated ground eigenstate for eleven topologies up to $N=6$ for the band $m=0$.  Eigenstates for higher bands $m$ are determined by replacing $|g,0\rangle$ by $|-,m\rangle$ and $|-,1\rangle$ by $|-,m+1\rangle$ in every instance.  The first four rows are valid for any topology.}
  \label{tab:states}
\end{table*}

The differences between topologies are apparent when comparing, for
example, the band $m=0$, for $2\leq\langle\hat{L}\rangle\leq N-1$, two
excitations in a square topology compared with two excitations in a
tetrahedron topology.  In the square case, there is a different
coefficient for excitations adjacent as to excitations separated,
compared with the tetrahedron case, where all terms have the same
coefficient.

Three different topologies (pentagon, six cavities with $z=3$, and six
cavities with $z=4$) have some coefficients displayed to three decimal
places, these have been calculated to twelve decimal places.  An exact
form is not derived, as these numbers represent solutions of
polynomials of order $\geq 50$ no exact form is necessarily expected.

We also examine the expectation value of the number of excitations of
each cavity, $\langle\hat{L}_i\rangle$ and
$\langle\hat{L}_i^2\rangle$, and find that, independent of topology,

\begin{equation}
\begin{split}
\label{eq:Li}
\langle \hat{L}_i \rangle =& \frac{\langle\hat{L}\rangle}{N} \\
\langle \hat{L}_i^2 \rangle =& (m+1)^2 \left(\frac{\langle\hat{L}\rangle-m N}{N}\right) \\
&+ m^2 \left(1-\frac{\langle\hat{L}\rangle - m N}{N}\right),
\end{split}
\end{equation}
so that the variance of $\hat{L}_i$ is

\begin{equation}
  \begin{split}
\label{eq:Li2}
    {\rm var}&(\hat{L}_i) = \sqrt{\langle\hat{L}_i^2\rangle - \langle \hat{L}_i\rangle^2} \\
&= \sqrt{\frac{\langle\hat{L}\rangle-m N}{N} - \left(\frac{\langle\hat{L}\rangle-m N}{N}\right)^2},
\end{split}
\end{equation}
where the band $m$ is defined by Eq.~(\ref{eq:band}), and both
Eqs.~(\ref{eq:Li}) and (\ref{eq:Li2}) are valid for $i=1,\ldots,N$.
From this, one can determine that if $\langle\hat{L}\rangle$ is an
integer multiple of $N$, ${\rm var}(\hat{L}_i)=0$, as expected.  Also,
if we consider the thermodynamic limit, where both the number of
cavities and the number of excitations approach infinity
($N\rightarrow\infty, \langle\hat{L}\rangle\rightarrow\infty$), while
the excitation density remains constant at $\rho =
\langle\hat{L}\rangle/N$, we find that ${\rm
  var}(\hat{L}_i)\rightarrow0$.

While this paper primarily focuses on phase changes as a function of
$\kappa$, one can also examine phase changes as a function of detuning
$\Delta$ \cite{ref:angelakis07}.  Indeed, experimentally shifts in
$\Delta$ may prove to be more accessible (via the Stark shift), as
$\kappa$ cannot be changed post-fabrication in many systems.  In
Fig.~\ref{fig:ofDelta}(a) we plot $\langle \hat{L}\rangle$ as a
function of $\Delta/\beta$ and $(\mu-\omega)/\beta$ for $\kappa=0$,
and in Fig.~\ref{fig:ofDelta}(b) we do the same for
$\kappa=10^{-1/2}\beta$.  There are fewer plateaus in (a) than there
are in (b) because there are fewer plateaus (due to the pinching
effect, discussed above) at $\kappa/\beta=0$.  Note the
\emph{symmetry} around $\Delta=0$, in the second and subsequent
boundaries of Fig.~\ref{fig:ofDelta}(a) (c.f.  Fig.~3 of
\cite{ref:GTCH}), and the corresponding \emph{asymmetry} in the third
and subsequent boundaries of Fig.~\ref{fig:ofDelta}(b).  This symmetry
is perfect at $\kappa=0$, and the asymmetry increases with increasing
$\kappa$.

\placeofDelta

A mean-field phase diagram that is comparable with the phase diagrams
of the previous section can be made \cite{ref:GTCH}.  An accurate
comparison between the exact results of the previous section, with
mean-field, is made when we consider topologies with $z$ nearest
neighbors with mean-field results for $z$ nearest neighbors.  In all
eleven distinct topologies tested, a very accurate match is seen.  One
such result is displayed in Fig.~\ref{fig:matches}.  We find that the
boundaries from Eq.~(\ref{eq:criticalchempot}) are preserved in the
mean-field solutions.

\placematches

In mean-field, the region with $\psi=0$ corresponds to the various
Mott insulating lobes (e.g., $|g,0\rangle$, $|-,1\rangle$,
$|-,2\rangle$ etc.), while $\psi>0$ is the superfluid state.  The
bottom lobe is described as the zeroth lobe ($|g,0\rangle$), the next
lobe up as the first lobe ($|-,1\rangle$), and so on.  In
Fig.~\ref{fig:superfill}(a), we examine the underside of the first
lobe in mean-field with $z=2$, and overlay the boundary between
$\langle\hat{L}\rangle=N-1$ and $\langle\hat{L}\rangle=N$ for
$N=2,3,4,5,$ and 6 cavities in periodic boundary conditions.  In
Fig.~\ref{fig:superfill}(b), we examine the underside of the second
lobe in mean-field with $z=2$, and plot the boundary between
$\langle\hat{L}\rangle=2N-1$ and $\langle\hat{L}\rangle=2N$ for
$N=2,3,4,5$ cavities, also in periodic boundary conditions.  One can
clearly see how as the number of cavities increases, the boundaries
approach the boundary of mean-field, and eventually may pinch off for
each lobe entirely (so that plateaus of height
$\langle\hat{L}\rangle/N = 1,2,\ldots$ do not continue as
$\kappa\rightarrow\infty$, but rather have finite size in this
direction).  These boundaries accord well with the structures observed
by Rosario and Fazio (Fig.~3 or Ref.~\cite{ref:pointylobes}), which
were obtained independently by the density matrix renormalization
group procedure, which lends weight to both quantum treatments.
Furthermore, as $N\rightarrow\infty$, our exact results approach that
of the mean-field, which have a more rounded cutoff for the Mott lobes
than these finite cavity results.

\placesuperfill

\section{Disorder and effective model temperature}
\label{sec:disordertemperature}

In this section, we first consider the modification of the chemical
potential with small temperature increase (less than the scale for
photon generation, $kT\ll\hbar\omega$) and hence this modifies the
phase diagrams above.  We then examine fabrication disorder in the
form of a normal distribution of photon energies for each cavity.  We
show that this fabrication disorder is qualitatively similar to
effective temperature, providing a connection between disorder and an
effective temperature in this analogue system.

Note that we set Boltzmann's constant $k_B=1$.  We begin by
differentiating the free energy $F=E-TS$ with respect to the total
number of excitations $l$,

\begin{equation}
  \frac{\partial F}{\partial l} = \frac{\partial E}{\partial l} - T \frac{\partial S}{\partial l} - S \frac{\partial T}{\partial l},
\end{equation}
recalling the definition of chemical potential in
Eq.~(\ref{eq:mudef}), and assuming that temperature does not depend on
the number of excitations (i.e.,~$\partial T/\partial l=0$, assuming
that the temperature scale is too low to generate a photon,
i.e.,~$kT\ll \hbar\omega$), we get

\begin{equation}
\mu = \frac{\partial E}{\partial l} + T\frac{\partial S}{\partial l},
\end{equation}
this then gives an effective chemical potential

\begin{equation}
\mu' = \mu + T \frac{\partial S}{\partial l}.
\end{equation}
We calculate the entropy $S$ in the $\kappa\rightarrow0$ limit in the
following manner.  Assume that the photon blockade is complete,
i.e.,~$|-,2,g,0\rangle \nrightarrow |-,1,-,1\rangle$, then we can
consider each band [recall Eq.~(\ref{eq:band})] separately.  More
specifically, each band acts like a paramagnet \cite{ref:amitverbin}.
Recall that a one-dimensional paramagnet is a line of spin states,
where each spin can point up or down.  Compare with our system, where
each cavity can have either $|g,0\rangle$, or $|-,1\rangle$ (for band
$m=0$).  Strictly speaking, each state (as in Table \ref{tab:states})
is a pure state, and as such the entropy is zero.  However, if we
assume that the number of cavities $N$ is very large, then the
superposition states acts like a mixed state, and we can define
entropy as for a paramagnet (essentially the logarithm of the number
of microstates) by

\begin{equation}
  S(l) = \ln{ N \choose l-m N}. 
\end{equation}
Note that this solution is only valid within each band, as such we can
ignore the infinities that arise in $\partial S/\partial l$ when $l$
is an integer multiple of $N$, as at these points the paramagnetic
approximation does not apply. Recall that the phase diagram of
$\langle \hat{L}\rangle$ is concerned with finding the slope of the
smallest energy eigenvalue with respect to $\mu$, and that this slope
is always an integer.  The Hamiltonian of Eq.~(\ref{eq:mainham}) is
block diagonal; we know from earlier analysis that the ground state
energy of each block $\Hl$ has constant slope with respect to $\mu$ of
$-l$.  Consider Fig.~\ref{fig:energies}, when temperature is included,
each line will move, with respect to $(\mu-\omega)/\beta$ by some
amount to the left or to the right.  For each $\mu$, we choose the
smallest energy eigenvalue at that point, and take the negative slope
at that point.  For small finite temperatures, this is manifest as a
``splitting'' of the pinches, as seen in \cite{ref:fisher}.  We plot
this splitting between the zeroth and first lobes, and the first and
second lobe, for $N=10,100,1000$ in
Fig.~\ref{fig:temperatureanddisorders}(a).

\placetemperatureanddisorders

Fabrication of a system of photonic cavities will undoubtedly be
subject to certain errors.  Here we model uncertainty in the cavity
frequency $\omega$.  We assume that each cavity may be tuned
individually to $\Delta_i=0 \; \forall i=1,\ldots,N$ (probably via
the Stark shift), and as such model the Hamiltonian by

\begin{equation}
\begin{split}
  \Ha = &\sum_{j=1}^N\Big[ (\omega_i + \delta_i) (\sigma^+_i\sigma^-_i + a^{\dagger}_i a_i ) \\
  & + \beta_i(\sigma^+_i a_i + \sigma^-_i a^{\dagger}_i)\Big] - \kappa
  \sum_{\langle i,j\rangle} a^{\dagger}_i a_j,
\end{split}
\end{equation}
where the set $\{\delta_1, \delta_2, \ldots, \delta_N\}$ is chosen
from a normal distribution with zero mean and standard deviation
$\varsigma$.  For fixed number of cavities and fixed $\varsigma$, we
calculate 1000 sets each of boundaries above $l=0$ to below $l=2N$,
and take the mean of the results.  Results are shown in
Fig.~\ref{fig:temperatureanddisorders}(b).

\placetstar

One can see by comparing Figs.~\ref{fig:temperatureanddisorders}(a)
and \ref{fig:temperatureanddisorders}(b) that disorder and temperature
produce qualitatively the same results.  However, these disorder
effects can only be calculated up to four cavities due to the
limitations of computing resources, and the temperature analysis is
only valid for large numbers of cavities.  Hence the two techniques
cannot be compared directly.  If the exact diagonalization technique
could be extended to a larger number of cavities, it could be compared
quantitatively with disorder, matching properly the standard deviation
with the effective temperature $T$.

If one envisions Fig.~\ref{fig:temperatureanddisorders}(a) as temperature
increases even further, there will be some temperature $T^*$ such that
the top line from the bottom group (corresponding to $l=N-2$) will
meet the bottom line of the top group (corresponding to $l=N+1$).  We
examine $T^*$ as a function of the number of cavities, and find that
this is given by

\begin{equation}
\begin{split}
T^* = (2-\sqrt{2}) \Big[ &\Gamma(N-2) +\Gamma(N+2)\\
-&\Gamma(N-1)-\Gamma(N+1)\Big]^{-1},
\end{split}
\end{equation}
where $\Gamma(l') = l' \left.\frac{\partial S}{\partial
    l}\right|_{l=l'}$, this function is plotted in
Fig.~\ref{fig:tstar}.  $T^*$ appears to converge to a constant,
non-zero temperature as $N\rightarrow\infty$.

\section{Conclusions}
\label{sec:conclusions}

In this paper we present an intensive analysis of the
Jaynes-Cummings-Hubbard model using the exact diagonalization
technique - studying the phase diagrams via the expectation value of
the total number of excitations.  We examine various topologies of
small networks of cavities, and compare this work with the infinite
cavity mean-field approximation, we find good agreement in all
topologies.  We study the effective model temperature, and compare
this qualitatively with disorder in the photon energy of the exact
JCH, and also find good agreement.

\section{Acknowledgments}

The authors would like to acknowledge useful discussions with
Z.E.~Evans, A.M.~Stephens and C.-H.~Su.  A.D.G. and
L.C.L.H. acknowledge the Australian Research Council for financial
support (Projects No. DP0880466 and No. DP0770715, respectively).
This work was supported in part by the Australian Research Council,
the Australian Government and by the US National Security Agency, and
the U.S. Army Research Office under Contract No.  W911NF-04-1-0290.

\bibliographystyle{apsrev} 
\bibliography{papers}

\end{document}